\documentclass[twocolumn]{aastex701} 

\usepackage{amsmath}
\usepackage{gensymb}

\usepackage[caption=false]{subfig}

\shorttitle{Hot Jupiters' Isolation Is Not Unique to High-Eccentricity Tidal Migration}

\shortauthors{Radzom et al.}

\graphicspath{{./}{figures/}}
\begin{document}

\title{Hot Jupiters' Isolation Is Not Unique to High-Eccentricity Tidal Migration}

\author[0000-0002-0015-382X]{Brandon T. Radzom}
\affiliation{Department of Astronomy, Indiana University, 727 East 3rd Street, Bloomington, IN 47405-7105, USA}
\affiliation{Caltech/IPAC-NASA Exoplanet Science Institute, 1200 E. California Boulevard, MC 100-22, Pasadena, CA 91125, USA}
\email{bradzom@iu.edu}

\author[0000-0002-7846-6981]{Songhu Wang}
\affiliation{Department of Astronomy, Indiana University, 727 East 3rd Street, Bloomington, IN 47405-7105, USA}
\email{sw121@iu.edu}

\author[0000-0002-5668-243X]{Bonan Pu}
\affiliation{Independent Researcher, New York, NY 10001, USA}
\email{bonanpu@gmail.com}

\author[0000-0002-5181-0463]{Hareesh Gautham Bhaskar}
\affiliation{Department of Astronomy, Indiana University, 727 East 3rd Street, Bloomington, IN 47405-7105, USA}
\email{hbhaskar@iu.edu}

\author[0000-0002-7670-670X]{Malena Rice}
\affiliation{Department of Astronomy, Yale University, 219 Prospect Street, New Haven, CT 06511, USA}
\email{malena.rice@yale.edu}

\correspondingauthor{Brandon T. Radzom}
\email{bradzom@iu.edu}


\begin{abstract}

Conventionally, the observed isolation of hot Jupiters, marked by a paucity of nearby low-mass planetary companions, has been interpreted as evidence of high-eccentricity tidal migration for these close-in gas giants. This loneliness is in stark contrast with the compact configurations commonly observed for warm Jupiters, indicating a curious dichotomy in dynamical origins between these two classes of short-period giants. In this study, however, we adopt a unified quiescent framework for both giant populations wherein they emerge from the protoplanetary disk as the sole Jovian planet within a compact multi-super-Earth system. We use long-term numerical simulations to show that post-disk dynamical evolution will naturally result in an observed preferential isolation for hot Jupiters relative to warm Jupiters. Specifically, their companions achieve significantly larger period ratios and mutual inclinations, rendering them more difficult to detect --- especially via the transit method. Additionally, we find that this paradigm is consistent with the enigmatic population of longer-period hot Jupiters hosting interior companions on coplanar orbits. Another prediction of this model, best tested through high-precision Doppler campaigns, is the existence of a population of close-in ($P\lesssim 50\,\mathrm{days}$) but mutually inclined outer companions to hot Jupiters. 

\end{abstract} 

\section{Introduction} 
\label{sec:intro}
The longest-standing mystery in exoplanet astronomy pertains to the origins of short-period Jovian-class planets such as hot Jupiters (as comprehensively reviewed by \citealt{Dawson2018}). Proposed theories to explain the existence of these gas giants can be categorized as either dynamically quiescent or violent. Mechanisms labeled as ``dynamically quiescent'' include disk-driven migration, wherein the giant forms in the outer regions before moving inward through interactions with their disk \citep{Goldreich_Tremaine1979, Lin_Papa1986, Lin1996}, and in situ formation, wherein the giant forms at or proximate to its current orbit \citep{Lee2014, Boley_2016, Batygin2016, Bailey2018}. The ``dynamically violent'' mechanisms encompass a variety of high-eccentricity tidal migration pathways which can be triggered by planet-planet scattering \citep{Rasio_Ford1996, Weidenschilling1996, Chatterjee2008}, von Zeipel-Lidov-Kozai cycling \citep{Lidov1962, Kozai1962, Wu_2003, Fabrycky2007, Naoz2016_review, Ito2019}, or other secular interactions \citep{Wu_2011, Lithwick2011, Petrovich2015, Hamers2017}.  

Independently, these formation and evolution channels can each explain only a subset of the known observational characteristics of short-period gas giants \citep{Dawson2018}. One such property, currently challenging each origin model individually, is the observed preferential isolation of hot Jupiters ($P < 10 \, \mathrm{days}$) relative to their wider-orbiting ($10 \, \mathrm{days} \leq P\lesssim 300 \, \mathrm{days}$) giant counterparts, known as warm Jupiters (see \citealt{Spalding_2017} for an exception). Specifically, the earliest population studies with the \textit{Kepler} transit survey demonstrated that hot Jupiters were effectively devoid of any nearby low-mass planetary companions, with less than 1\% hosting a nearby co-transiting neighbor, while more than half of warm Jupiters had such companions (e.g., \citealt{Steffen2012, Huang_2016}). This stark companionship dichotomy served as strong evidence of dichotomous histories for these two populations of short-period giants, with quiescent mechanisms being dominant for warm Jupiters, and violent high-eccentricity migration, which is capable of largely clearing out any close-in companions \citep{Mustill2015,Becker_2026}, producing most hot Jupiters. However, the latest results from the Transiting Exoplanet Survey Satellite (TESS) mission and transit timing variation (TTV) searches for nearby companions over the complete \textit{Kepler} dataset firmly establish that a non-negligible fraction ($\gtrsim 7$--$12\%$) of hot Jupiters are not lonely \citep{Hord_2021, Sha2026, Wu_2023}, and hence may have shared origins with warm Jupiters.

In this work, we investigate whether a unified, dynamically quiescent framework for both hot Jupiters and warm Jupiters can explain their observed companionship dichotomy. Specifically, we use long-term $N$-body simulations of short-period giants in compact multi-planet systems to demonstrate that, through gravitational interactions alone, hot Jupiters naturally self-isolate relative to warm Jupiters. We detail the setup of our fiducial simulations in Section \ref{sec:simulation_setup}. In Section \ref{sec:fiducial_results}, we present the results of these simulations, characterizing the dichotomy that forms for the orbital spacings and mutual inclinations (Section \ref{subsec:isolation_10days}) as well as the observational detection efficiencies  (Section \ref{subsec:obs_dichotomy}) of our companion planets, and also highlighting the key properties of our final system architectures (Section \ref{subsec:architectures}). We then describe the dynamical evolution that drives these companionship outcomes in Section \ref{sec:dynamics}, and present results from simulation suites with modified properties in Section \ref{sec:nonfiducial_results}. We discuss our adopted quiescent model in the context of in situ formation and disk migration in Section \ref{sec:compatability}, and summarize our main findings and their potential implications in Section \ref{sec:conclusions}.

\section{Simulation Set-up}
\label{sec:simulation_setup}

\textbf{Overview.} We generate 360 primordial ``peas-in-a-pod'' \citep{Weiss2018,Millholland_Wang_laughlin2017,Goyal2022} planetary systems that resemble more compact, higher-multiplicity versions of \emph{Kepler} multiple-planet systems, which are packed just wide of the boundary of Hill stability \citep{Pu2015, Volk_2015}. Such a set-up is well-motivated by evidence showing that current-day \emph{Kepler} multi-planet systems may have undergone significant dynamical sculpting in the past, originating from pristine systems with more planets in initially tighter configurations (e.g., \citealt{Izidoro_2017, Zink_2019, Izidoro_2021}). Each of our systems begins with twelve nearly uniformly and closely spaced super-Earths with similar masses orbiting a solar-mass star (analogous to the set-up used in \citealt{Goldberg_2022}). Before integration, we select one super-Earth in each system to have a wider spacing relative to its nearest neighbors and grow to become a Jupiter-mass ($1\,\text{M}_{\mathrm J}$) giant, which will trigger dynamical instability in the absence of eccentricity damping from the disk \citep{Ward1997,Chatterjee2008}. This numerical experiment is designed primarily to explore long-term dynamical evolution as a mechanism to produce a companionship dichotomy between hot and warm Jupiters under unified formation conditions. More realistic initial conditions for short-period giants in such compact, high-multiplicity systems are likely more complex (e.g., 
see \citealt{Ogihara2013,Ogihara2014}).

\textbf{Planetary Mass \& Radius.} We assign planetary masses according to a Gaussian distribution centered on $6\,\text{M}_\oplus$ (\citealt{Wu_2019}; consistent with typical masses in \emph{Kepler} multi-planet systems) with $\sigma_m= 20\%\times6\,\text{M}_\oplus =1.2\,\text{M}_\oplus$ and within the range $1\,\text{M}_\oplus-10\,\text{M}_\oplus$. We determine planetary radii from their masses using the mass-radius relation for rocky, volatile-poor bodies described in Equation 1 of \cite{Otegi2020}. We note that our results do not sensitively depend on the exact masses (nor radii) of our small planets, as they are much less massive than our giants.

\textbf{Spacing.} To generate probable progenitors to \emph{Kepler}-like peas-in-a-pod systems, we employ a prescription similar to the Equal Mutual Separation (EMS) scheme (e.g., \citealt{CHAMBERS1996}; \citealt{Zhou2007}). We set the planet-planet spacing parameter to $K=10$, which corresponds to the critical boundary for instability defined by \cite{Funk2010} for $N\geq8$ systems as well as the compact edge of the expected distribution for observed \emph{Kepler} multi-planet systems \citep{Pu2015}, and is broadly consistent with expectations from planet formation (e.g., \citealt{Kokubo1998}). The spacing parameter plays an important role in setting the dynamical stability timescale for EMS systems \citep{OBERTAS2017} and is defined as
\begin{equation}
    K = \frac{a_j-a_{j-1}}{r_{H,\mathrm{mut}}}
\end{equation}
where $a_j$ and $a_{j-1}$ are the semi-major axes of the $j^{th}$ and $(j-1)^{th}$ planets from the star, and $r_{H,\mathrm{mut}}$ is their mutual Hill radius
\begin{equation}
    r_{H,\mathrm{mut}} = \frac{a_j+a_{j-1}}{2} \left ( \frac{m_j+m_{j-1}}{3 M_*} \right )^{1/3}
\end{equation}
where $m_j$ and $m_{j-1}$ are the masses of the $j^{th}$ and $(j-1)^{th}$ planets, respectively, and $M_*$ is the mass of the central star.

Since the planetary masses within each of our simulated near-EMS systems are similar, the period ratios $P_\mathrm{out}/P_\mathrm{in}$ between adjacent planets are also nearly identical, with a characteristic value of $1.43$--- intermediate of the 7:5 and 3:2 mean motion resonances (MMRs). To sample a more continuous distribution for the orbital periods of our giants, we also vary the period of the innermost planet by drawing from a half-normal distribution with mean $\mu_P=3\,\mathrm{days}$ (near the observed inner edge of multi-planet systems; \citealt{Mulders2018}) and $\sigma_P=10\% \times \mu_P=0.3 \,\mathrm{days}$. The minimum possible innermost period is thus set at $\mu_P$, and we impose a maximum value for this period at $4\,\mathrm{days}$. Given our spacing scheme and selected multiplicity ($N=12$), our simulations probe a period range of $3 \, \mathrm{days} \leq P\lesssim 300 \, \mathrm{days}$, which is consistent with the finite range identified for observed peas-in-a-pod systems \citep{Millholland_2022}. We note that the exploration of resonant configurations is beyond the scope of this work and discuss potential implications for companions initially in MMRs in the context of disk migration in Section \ref{sec:compatability}. 

\textbf{Eccentricity \& Inclination.}
In agreement with known \emph{Kepler} compact multi-planet systems, we ensure that the planets in our systems begin on orbits that are nearly circular and coplanar \citep{Xie_2016, Zhu_2018, Millholland_2021}. We limit the orbital eccentricities $e$ of each planet to be less than the orbit-crossing value $e_\text{c}$ of each system, computed as
\begin{equation}
\label{eq:ecross}
    e_\text{c} \approx \frac{1}{2} \left [ \left ( \frac{a_\text{outer}}{a_\text{inner}} \right )^{1/(N-1)} -1 \right ]
\end{equation}
where $N$ is the total number of planets in the system (always initially 12), and $a_\text{outer}$ and $a_\text{inner}$ are the outermost and innermost semi-major axes of each planetary system, respectively. The typical value expected for each system in our fiducial simulations is $e_\text{c}=0.15$. With this upper limit established, we pull eccentricities for each planet from a Rayleigh distribution with standard deviation $\sigma_e = 10\% \times e_\text{c}= 0.015$. We draw orbital inclinations from a Rayleigh distribution as well with $\sigma_{\theta} =  \sigma_e/2 \times (180 \degree / \pi)$, and impose an upper limit of $3 \degree $. 

\textbf{Oscillating Angles.} The values for the longitude of the ascending node ($\Omega$), as well as the sum of the argument of periapsis ($\omega$) and the true anomaly ($M$), are chosen randomly from $[0,2 \pi]$. 

\textbf{Giant Planet Formation.} With our peas-in-a-pod systems in place, we then assume that one super-Earth core in each system grows to become a giant planet. While the details of the disk properties and physical mechanisms required for such a planet to achieve critical mass and undergo runaway accretion are uncertain, we attempt to account for this uncertainty by increasing the mutual spacing of the ``proto-giant'' with its nearest neighbors by a random amount. Specifically, we determine this spacing enhancement $\delta K$ by pulling from a Gaussian distribution with a large standard deviation $\sigma_K=10$, such that the typical $K$ for this planet is double its initial value. An upper limit is imposed on $\delta K$ such that the final modified spacing, after the selected planet has been boosted to become a giant, cannot exceed the fiducial $K=10$ spacing. This local spacing buffer is consistent with a range of possible scenarios wherein the proto-giant super-Earth core opens a partial gap in the disk \citep{Rafikov2002,Paardekooper2004,Paardekooper2006,Ogihara2014}, potentially accreting materials from its neighboring planets' reservoirs. 

Once the chosen super-Earth's spacing is enhanced, we boost its mass to $1\,\text{M}_{\mathrm J}$ and radius to $1 \,\text{R}_{\mathrm J}$. We sample each of the twelve possible spots for giant insertion in each system (from $j=1$--$12$) 30 times. Since every super-Earth progenitor system is generated from scratch, we end up with $12\times30=360$ unique compact multi-planet systems in total, each containing one giant. The emergence of a Jovian-class planet generally reduces the minimum $K$ value in our systems, of which the instability timescale is a strong function \citep{Wu_2019}. The full range of minimum spacings in our systems is $3\lesssim K\leq10$, with a median value of $K\approx5$. Therefore, while most minimum spacings sit safely above the $K=3.5$ limit for two-body Hill instability \citep{Gladman_1993}, many also fall below the instability limits needed for high-multiplicity ($N>3$) systems (\citealt{Funk2010,OBERTAS2017}), ensuring dynamical disruption ensues in our multi-planet systems.

\textbf{Integration.} We utilize the hybrid integrator \texttt{TRACE} in \texttt{rebound} \citep{rebound, reboundias15, reboundtrace} to run the $N$-body integrations for each of our 360 systems. As the dominant body in each system, the giant planet's orbital period $P_\mathrm{giant}$ represents an important dynamical timescale in these simulations. In order to balance achieving equivalent dynamical ages and computational cost, we integrate each system to the lesser of either: 1) $10^9$ giant orbits (i.e., $10^9 P_\mathrm{giant}$) or 2)  $20\,\mathrm{Myr}$. Direct physical collisions are enabled and resolved with the conservation of mass and momentum. Planets are considered ejected and are removed from the system once their distance from the central star is greater than 250 AU. Commentary on conserved quantities and other relevant integration details is provided in Appendix \ref{app:energy_AM}.

\section{Fiducial Simulation Results}
\label{sec:fiducial_results}

\subsection{Hot Jupiters' Companions Are More Distant and Inclined System-Wide}
\label{subsec:isolation_10days}
\begin{figure*}
    \centering
    \includegraphics[width=0.9\textwidth,scale=1.0]{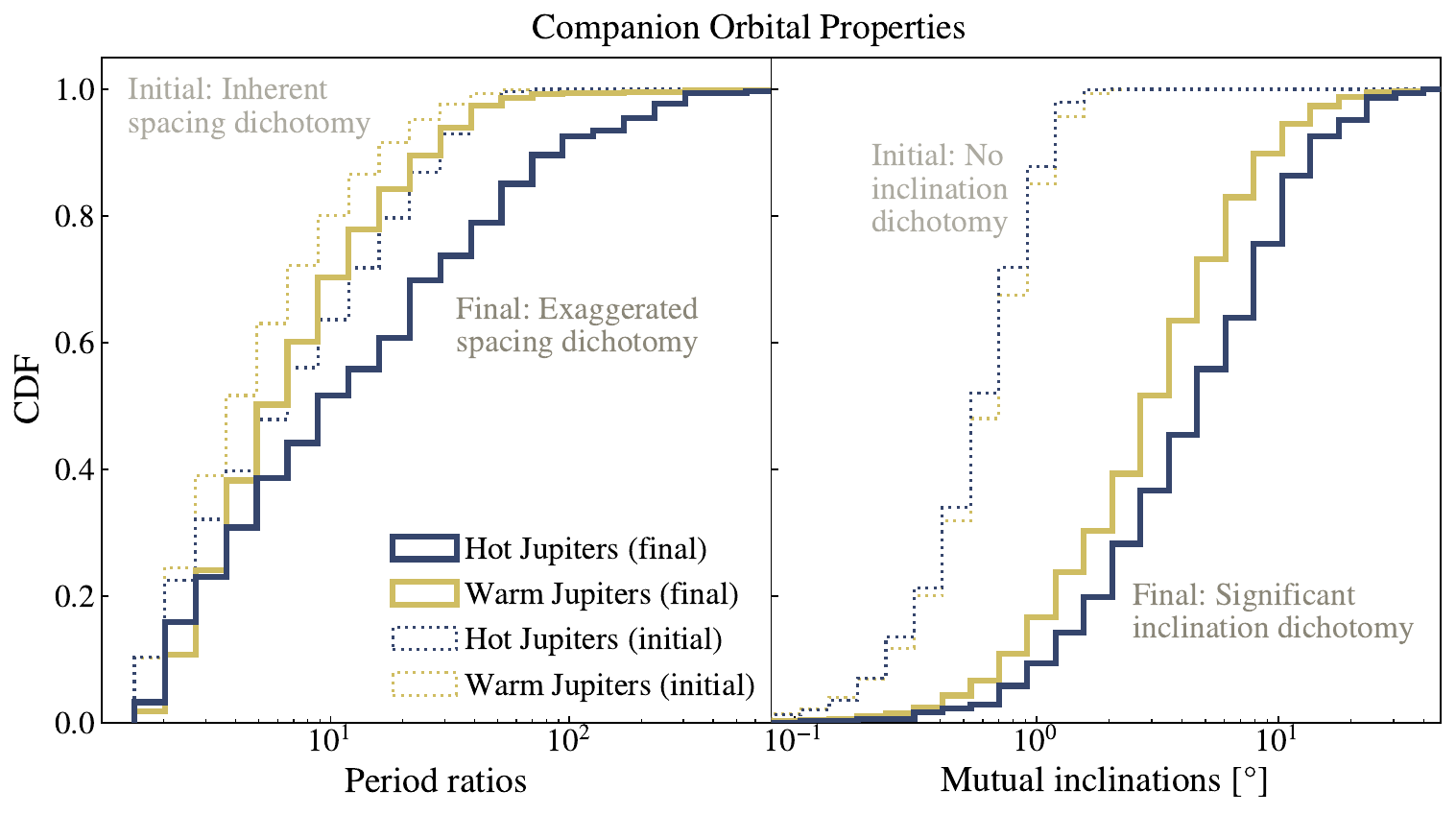}
    \caption{CDFs for the period ratios (left) and mutual inclinations (right) of our giant planets and their super-Earth companions, plotted separately for hot Jupiter systems (blue) and warm Jupiter systems (yellow) at the start (dotted lines) and end (solid lines) of our fiducial simulations. Hot Jupiters' companions begin with larger period ratios than warm Jupiters', but this discrepancy grows through dynamical evolution, which also preferentially excites the mutual inclinations of hot Jupiters' companions.}
    \label{fig:PR_RMSImut_CDFs}
\end{figure*}

\textbf{Period Ratio Dichotomy.} Each of our short-period giants retains at least one planetary companion at the end of the simulation, and we find similar mean companion multiplicities of $3.3$ and $3.4$ for our hot Jupiters ($P_\mathrm{giant}<10\,\mathrm{days}$) and warm Jupiters ($P_\mathrm{giant}\geq10\,\mathrm{days}$), respectively. To quantify the relative degree of isolation for both giant populations, we compute their period ratios with respect to each remaining companion. We find that hot Jupiters boast a $\approx70\%$ larger median value ($\mathrm{median}(P_\mathrm{out}/P_\mathrm{in})=11.0^{+3.1}_{-0.5}$) compared to warm Jupiters ($\mathrm{median}(P_\mathrm{out}/P_\mathrm{in})=6.5^{+0.5}_{-0.1}$). To measure the statistical significance of this difference, we compute the absolute value of the $z$-score associated with these median estimates, adopting the appropriate asymmetric upper and lower errors. We find $|z\mathrm{-score}|=6.6$, indicating a significant discrepancy (i.e., a $6.6\sigma$ enhancement for hot Jupiters' companions if the period ratios were Gaussian-distributed). In the left panel of Figure \ref{fig:PR_RMSImut_CDFs}, we plot the cumulative distribution function (CDF) of these period ratios for hot Jupiters and warm Jupiters separately, demonstrating a clear dichotomy between the two giant populations. Driven by the subset of hot Jupiter companions that sit at the wide end of CDF ($P_\mathrm{out}/P_\mathrm{in}\gtrsim10^2$), we find that the mean period ratio for these giants ($\mathrm{mean}({P_\mathrm{out}/P_\mathrm{in}})=47.6$) is $\approx2\times$ larger than for warm Jupiters ($\mathrm{mean}({P_\mathrm{out}/P_\mathrm{in}})=23.6$).

As another statistical measure of this period ratio dichotomy, we perform a two-sample Anderson-Darling (AD) test \citep{ADtest_Darling1957}, which is similar in nature to the well-known Kolmogorov–Smirnov (KS) test \citep{Kolmogorov_Morrison_1950, Smirnov1939}, but has been demonstrated to have superior statistical power \citep{Scholz_Stephens1987, Hou2009, Engmann_Cousineau2011}. To ensure the precision in our computed $p$-value lies well below commonly-adopted significance thresholds (e.g., $0.05$ or $0.005$; \citealt{Wasserstein2016}), we utilize the permutation method (implemented in \texttt{scipy.stats.anderson\_ksamp}) with $10,000$ re-samplings rather than rely on interpolation of a grid of pre-calculated values \citep{Scholz_Stephens1987, FeigelsonBabu2013}, which allows us to measure down to $p=10^{-4}$ (see, e.g., \citealt{Goyal2025}). The null hypothesis we test is that the underlying distributions of companion period ratios are the same for our simulated hot Jupiters (308 companions) and warm Jupiters (903 companions). We find this null hypothesis is firmly rejected by the AD test, and compute a resulting $p$-value at our precision floor of $p\leq10^{-4}$.

\textbf{Mutual Inclination Dichotomy.} In addition to becoming globally more widely-spaced, we find that the companions of hot Jupiters are also driven to orbits with higher mutual inclinations $\mathcal{I}_\mathrm{mut}$ relative to the plane of the giant\footnote{To calculate mutual inclinations, we first compute the complex inclinations $\mathcal{I}$ of the companion and the giant as $\mathcal{I} = \theta e^{i\Omega}$ where $i$ is the imaginary unit, and $\theta$ and $\Omega$ are the orbital inclination and longitude of the ascending node, respectively. Then, we calculate the absolute difference between these complex values to get the mutual inclination at the final timestep. Finally, to account for potentially significant inclination oscillations, we compute these inclinations over 500 orbits and adopt the root-mean-square as the mutual inclination $\mathcal{I}_\mathrm{mut}$.}. Specifically, we compute $\mathrm{median}(\mathcal{I}_\mathrm{mut})=5.3^{+0.4}_{-0.2}$$ \degree$ and $\mathrm{mean}({\mathcal{I}_\mathrm{mut}})=7.5\degree$ for our hot Jupiter systems, and $\mathrm{median}(\mathcal{I}_\mathrm{mut})= 3.4^{+0.2}_{-0.1}$$\degree$ and $\mathrm{mean}({\mathcal{I}_\mathrm{mut}})=4.9\degree$ for our warm Jupiter systems, corresponding to a $\approx55\%$ enhancement for hot Jupiters' companions and a significant $|z\mathrm{-score}|$ of $7.0$. The distributions of companion mutual inclinations for both giant populations are also statistically distinct as per AD test, which yields a $p$-value of $\leq10^{-4}$ (see also the right panel of Figure \ref{fig:PR_RMSImut_CDFs}). We summarize the results from this subsection in Table \ref{tab:results_summary}. Given that the conventional hot Jupiter-warm Jupiter boundary at $P=10\,\mathrm{days}$ is not astrophysically motivated, we also verify that the significance of the period ratio and mutual inclination dichotomy between the two simulated populations is robust to the precise boundary adopted.

\begin{deluxetable}{cccc}
\tabletypesize{\scriptsize}
\tablecaption{Summary of Key Companionship Results for Fiducial Simulations}\label{tab:results_summary}
\tablehead{\colhead{Metric$^a$} &
\colhead{Hot Jupiters} & \colhead{Warm Jupiters} & \colhead{$|z\mathrm{-score}|$}}
\tablewidth{300pt}
\startdata
\multicolumn{3}{l}{Period Ratio}\\ \\
    median($P_\mathrm{out}/P_\mathrm{in}$) & $11.0^{+3.1}_{-0.5}$ & $6.5^{+0.5}_{-0.1}$ & $6.6$\\
    mean($P_\mathrm{out}/P_\mathrm{in}$) & $47.6$ & $23.6$ & \nodata 
    \\\\ \hline
    \multicolumn{3}{l}{Mutual Inclination}\\ \\
    median($\mathcal{I}_\mathrm{mut}$) & $5.3^{+0.4}_{-0.2}$$\degree$ & $3.4^{+0.2}_{-0.1}$$\degree$ & $7.0$
    \\
    mean($\mathcal{I}_\mathrm{mut}$) & $7.5\degree$ & $4.9\degree$ & \nodata
    \\\\ \hline
    \multicolumn{3}{l}{Detectability}\\ \\
     $f(p_\mathrm{T}>0.5)$ & $16.6\pm2.1\%$ & $35.2\pm1.6\%$ & $7.1$ \\
    $f(V_\mathrm{TTV}>30\,\mathrm{min})$ & $9.6\pm3.0\%$ & $33.1\pm2.8\%$ & $5.6$ \\
    $f(K_\mathrm{RV}>3\,\mathrm{m/s})$ & $43.8\pm2.8\%$ & $55.5\pm1.7\%$ & $3.6$ \\\\ 
\enddata
\tablenotetext{a}{Reported uncertainties are standard errors; for median values, these are calculated using the 84th and 16th percentiles, normalized by the number of companions.}
\end{deluxetable}

\subsubsection{Inherent vs. Evolved Isolation}
\label{subsubsec:inherent_vs_evolved}
It is notable that the hot Jupiters in our simulations are born with fewer nearby companions than warm Jupiters and thus begin with a companion period ratio distribution that is inherently skewed to larger values. This is a natural consequence of the EMS scheme adopted in this work (see Section \ref{sec:simulation_setup}); the initial companion period ratios for giants occupying the $n$th innermost position in their 12-planet system (at position $j$) mirrors those for giants in the $n$th outermost position (at position $13-j$). Therefore, giants placed near the edges of their systems begin with period ratio distributions that exhibit more variance and a higher overall median than those embedded near the center, whose distributions are comparatively concentrated at lower values. As such, any period cut that breaks this symmetry will result in samples with disparate initial period ratio distributions. In this way, our hot Jupiters at $<10\,\mathrm{days}$, which initially occupy only the innermost four positions ($j=1$--$4$) in their systems, are predisposed to end up with more sparsely spaced companions than warm Jupiters, which jointly contain giants positioned in both the outermost four spots ($j=9$--$12$) as well as the central four spots ($j=5$--$8$) of their systems.

We characterize this inherent isolation effect by comparing the companion period ratios at the end of the simulation with their starting values, and include the initial CDFs in Figure \ref{fig:PR_RMSImut_CDFs}. As illustrated, our hot Jupiters indeed begin preferentially isolated relative to warm Jupiters. However, this inherent dichotomy is clearly exaggerated significantly through the dynamical evolution process. In particular, we find that the difference between the median (mean) period ratios of hot and warm Jupiters' companions increases by $2\times$ ($\approx5\times$) by the end of the simulations. Therefore, our initial conditions contribute to but do not fully account for the resultant companion spacing dichotomy between hot and warm Jupiters presented in Section \ref{subsec:isolation_10days}. We also plot the CDFs of the initial companion mutual inclinations for both giant populations in Figure \ref{fig:PR_RMSImut_CDFs}, which are statistically indistinguishable (each corresponding to median and mean values of $\sim 0.7\degree$), but evolve to become dichotomous.

\subsection{Hot Jupiter Companion Architectures} 
\label{subsec:architectures}
\begin{figure*}
    \centering
    \includegraphics[width=0.9\textwidth, scale=1]{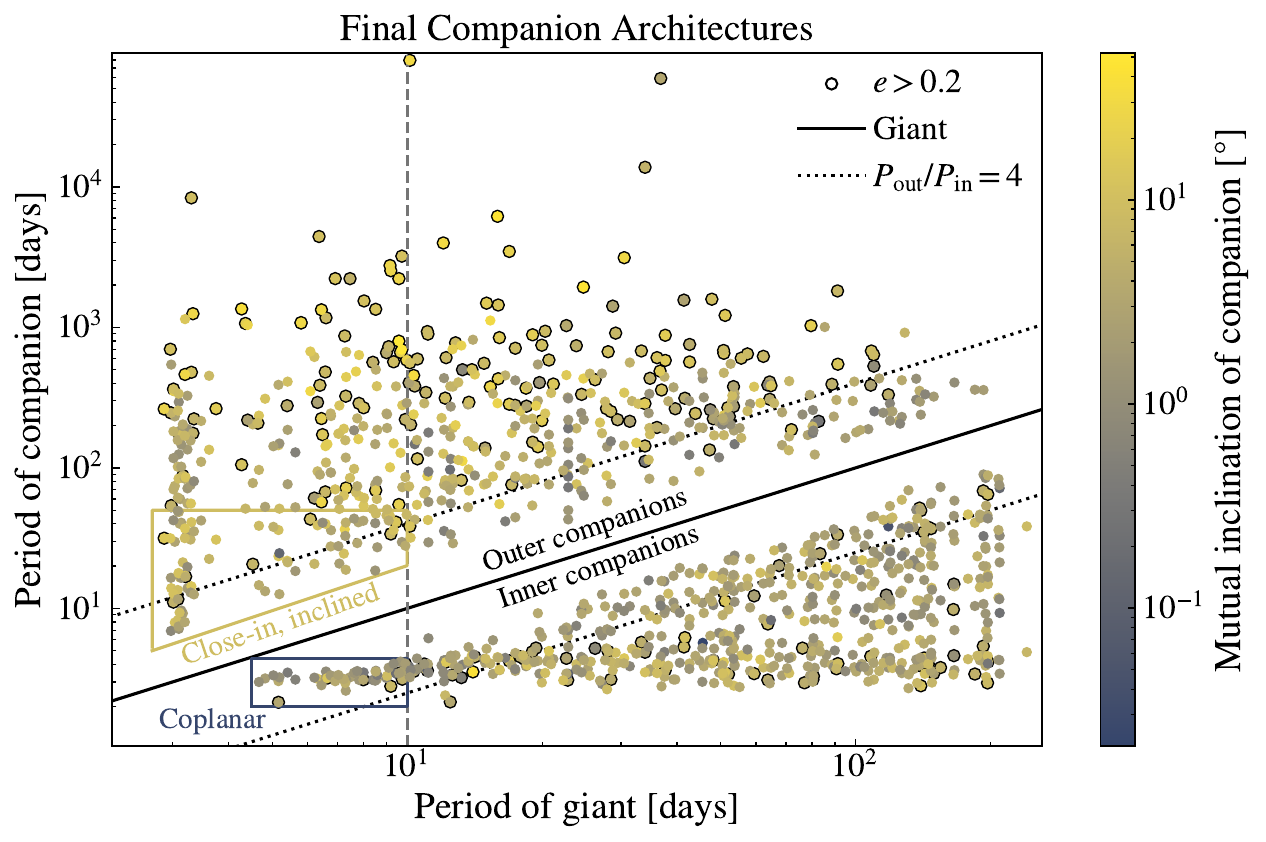}
    \caption{Final orbital periods of all low-mass planetary companions as a function of giant period, with companion-giant mutual inclinations represented on the color axis. The location of the giant is represented by the solid 1:1 line, above and below which all outer and inner companions lie, respectively, and the diagonal dotted lines correspond to a constant period ratio of 4. The vertical dashed line corresponds to the hot Jupiter-warm Jupiter boundary at $P=10\,\mathrm{days}$, and black circles indicate companions with eccentric orbits ($e>0.2$). We highlight two notable populations of nearby companions for hot Jupiters: 1) interior companions on co-planar orbits (blue rectangular region) and 2) exterior, close-in ($P<50\,\mathrm{days}$) companions with large mutual inclinations (yellow trapezoidal region).}
    \label{fig:Pfgiant_Pfcomp}
\end{figure*}

To visualize the architectures of our final systems, we plot the final orbital periods of our giants against their companions' periods in Figure \ref{fig:Pfgiant_Pfcomp}, with mutual inclinations represented on the color axis and eccentric companions ($e>0.2$) highlighted. We see that both hot and warm Jupiters frequently host nearby companions, which may take on either interior or exterior orbits. While warm Jupiters' inner companions have a relatively broad, uniform orbital period distribution, there is a clear concentration near $P\approx3\,\mathrm{days}$ for hot Jupiters' inner companions. Additionally, it is apparent that these interior companions have significantly lower mutual inclinations relative to the bulk of the companion population--- especially compared to hot Jupiters' outer companions. We discuss the properties of hot Jupiters' inner and outer companions in more detail below. 

\subsubsection{Coplanar Inner Companions to Hot Jupiters}
\label{subsubsec:sweet_spot}
The relatively flat period distribution for inner companions clustered near $\approx 3\,\mathrm{days}$ and spanning giant periods of $\approx4.5$--$10\,\mathrm{days}$ in Figure \ref{fig:Pfgiant_Pfcomp} demonstrates that hot Jupiters, particularly those with longer periods, are able to retain their innermost companion. Specifically, we find that 82\% of hot Jupiters with $P_\mathrm{giant}=4.5$--$10\,\mathrm{days}$ host an inner companion, compared to just 53\% of the full hot Jupiter population. Notably, these surviving interior companions tend to maintain low mutual inclinations ($\mathrm{median}(\mathcal{I}_\mathrm{mut})=1.7\degree$; $\mathrm{mean}({\mathcal{I}_\mathrm{mut}})=2.6\degree$), especially relative to hot Jupiters' outer companions ($\mathrm{median}(\mathcal{I}_\mathrm{mut})=6.4\degree$; $\mathrm{mean}({\mathcal{I}_\mathrm{mut}})=8.4\degree$). These inner companions, highlighted via the blue boxed region in Figure \ref{fig:Pfgiant_Pfcomp}, also have somewhat less eccentric orbits ($\mathrm{median}(e)=0.10$; $\mathrm{mean}({e})=0.12$) compared to the outer companions ($\mathrm{median}(e)=0.13$; $\mathrm{mean}({e})=0.17$).

We find that observed systems show good qualitative agreement with this feature. As of this writing, ten of eleven hot Jupiters known to reside in compact configurations with small planets host a nearby companion on a co-transiting (i.e., likely coplanar), interior orbit (\citealt{Becker_2015,Canas_2019,Huang_2020,Hord_2022,Maciejewski2023,Sha_2023,Korth2024,McKee_2025, Quinn2026}; see \citealt{Hord2024} for the exception). While this is consistent with geometric transit bias, it is intriguing that these giants with inner companions also tend to occupy the long tail of the orbital period distribution for hot Jupiters (see e.g., \citealt{Santerne2016, Yee2023}), spanning from $ 4.2\,\mathrm{days}$ (WASP-47\,b; \citealt{Hellier_2012}) to $ 8.5\,\mathrm{days}$ (WASP-84\,b; \citealt{Anderson2014})--- a trend borne out by our simulations. In most cases, the data and/or modeling for these observed systems are insufficient to determine whether these companions have significantly non-zero eccentricity (e.g., WASP-84\,c; \citealt{Maciejewski2023}), often only enabling upper limits based on stability arguments (see e.g., the case of WASP-47; \citealt{Becker_2015} or TOI-1130; \citealt{Huang_2020}), so their underlying eccentricity distribution remains poorly constrained. 
 
\subsubsection{Inclined, Close-in Outer Companions to Hot Jupiters}
\label{subsubsec:outer_closein_HJcomps}
A feature that is ubiquitous to all our hot Jupiters is the presence of outer companions at relatively short orbital periods: we find that 30\%, 70\%, and 90\% of hot Jupiters host at least one outer companion with $P<20\,\mathrm{days}$, $<50\,\mathrm{days}$, and $<100\,\mathrm{days}$, respectively. It is also not uncommon for multiple outer companions to occupy this domain; e.g., the average multiplicity of exterior-orbiting companions within $P<100\,\mathrm{days}$ is $\approx1.5$.

Observationally, only two examples of close, low-mass outer companions to hot Jupiters are known, and both occupy very close-in, co-transiting orbits: WASP-47\,d at $P=9.03\,\mathrm{days}$ \citep{Becker_2015} and TOI-4468.02 at $P=7.01\,\mathrm{days}$ \citep{Hord2024}. While this is also relatively unsurprising given the diminishing transit probability at wider separations, we find this may also be explained by the elevated mutual inclinations inherent to our hot Jupiters' exterior-orbiting companions. This is especially true for those at longer periods: we calculate $\mathrm{median}(\mathcal{I}_\mathrm{mut})=[3.4, 4.6, 5.1]\degree$ ($\mathrm{mean}({\mathcal{I}_\mathrm{mut}})=[5.2, 6.0, 6.4]\degree$) for outer companions with $P<[20, 50, 100]\,\mathrm{days}$. These inclined close-in companions, roughly bounded by the yellow region in Figure \ref{fig:Pfgiant_Pfcomp}, are less likely to occupy the co-transiting plane with the giant and hence be detected via transit surveys. Tentative evidence for such hidden companions comes from the \textit{Kepler} TTV search conducted by \cite{Wu_2023}, which was partially sensitive to non-transiting companions.

Critically, we note that the population of inclined, close-in, outer companions to our hot Jupiters (highlighted in Figure \ref{fig:Pfgiant_Pfcomp}) boasts higher mutual inclinations than is assumed in most transit-based, completeness-corrected occurrence rate calculations. Indeed, hot Jupiter companionship studies based on \textit{Kepler} and TESS data adopt coplanar or near-coplanar (i.e., $\mathcal{I}_\mathrm{mut}\lesssim2\degree$) companion inclination distributions in their canonical nearby companion estimates (\citealt{Huang_2016,Hord_2021,Sha2026}). In the near-coplanar case, the common assumption for their mutual inclinations is a Rayleigh distribution with $\sigma=1.8\degree$ (see e.g., \citealt{Huang_2016,Sha2026}), motivated by typical multi-planet systems consisting of super-Earth and sub-Neptune-like planets \citep{Fabrycky2014}. With this prescription, it is expected that $\approx60\%$ of hot Jupiters' companions boast $\mathcal{I}_\mathrm{mut}>1.8\degree$--- in our simulations, this fraction is $\approx90\%$ for hot Jupiters' outer companions (and 82\% for all surviving companions). Thus, the true rate of nearby companions to hot Jupiters may be underestimated by transit surveys, particularly for calculations covering longer orbital period regimes ($\gtrsim20\,\mathrm{days}$) where we find companions are abundant but have significantly higher mutual inclinations. We explore this possibility further in Section \ref{subsubsec:transits}.

\subsection{An Observational Companionship Dichotomy}
\label{subsec:obs_dichotomy}
In order to better understand the observational implications of the companion period ratio and mutual inclination dichotomies between our hot Jupiter and warm Jupiter populations (Section \ref{subsec:isolation_10days}), we investigate the relative detectability of our companions through transit, TTV, and radial velocity (RV) techniques. Rather than attempt population synthesis and rigorously apply our simulated results to a specific survey's completeness or detection pipeline (e.g., \emph{Kepler} or TESS), we instead compare representative companion ``detection'' fractions $f(X)$ between our hot and warm Jupiter systems after applying some threshold $X$ for detection via each observational method. We use these detection fractions largely as a point of comparison between hot and warm Jupiter systems, not as absolute indicators of detectability. 
As part of our transit analysis, we also explicitly characterize companion detectability over the same period ranges scrutinized in relevant observational studies. 

\subsubsection{Transits}
\label{subsubsec:transits}
For the transit method, we consider the geometric transit probability of our giants' companions, utilizing a similar approach to the methodology employed in Appendix D of \cite{Wu_2023}. First, assuming a circular orbit, we randomly select a transiting orbital inclination for the giant planet over $\theta_\mathrm{giant} \in \mathcal{U}(90 \degree -\arctan{(R_*/a)},90 \degree +\arctan{(R_*/a)})$. Then, for each companion, we compute the associated inclination of the planet $\theta_\mathrm{comp}$ given their mutual inclination $\mathcal{I}_\mathrm{mut}$ and apply the following criterion: $90 \degree -\arctan{(\frac{R_*}{a(1-e^2)})} < \theta_\mathrm{comp} <90 \degree +\arctan{(\frac{R_*}{a(1-e^2)})}$, which takes into account the often non-negligible eccentricities of our small planets \citep{Burke2008}. We repeat this process 10,000 times to get a robust transit probability $p_\mathrm{T}$ for each companion.

We plot the CDF of all probabilities for hot Jupiters and warm Jupiters in the left panel of Figure \ref{fig:detectability_CDFs}. Adopting $p_\mathrm{T} >0.5$ (i.e., $>50\%$) as the threshold for detection (highlighted also in Figure \ref{fig:detectability_CDFs}), we calculate the the fraction of detectable companions $f(p_\mathrm{T}>0.5)$ around both giant populations as well. We find $f(p_\mathrm{T}>0.5)=16.6\pm2.1\%$ for hot Jupiters and $f(p_\mathrm{T}>0.5)=35.2\pm1.6\%$ for warm Jupiters, indicating that hot Jupiters' companions are significantly ($|z\mathrm{-score}|=7.1$) less likely to be co-transiting compared to warm Jupiters'. We report these results in Table \ref{tab:results_summary}. Following our methodology outlined in Section \ref{subsec:isolation_10days}, we also perform an AD test on the $f(p_\mathrm{T}>0.5)$ distributions for these two giant populations, verifying that they are statistically distinct ($p\leq10^{-4}$).

\begin{figure}[htb]
    \centering
    \includegraphics[width=0.47\textwidth,scale=0.75]{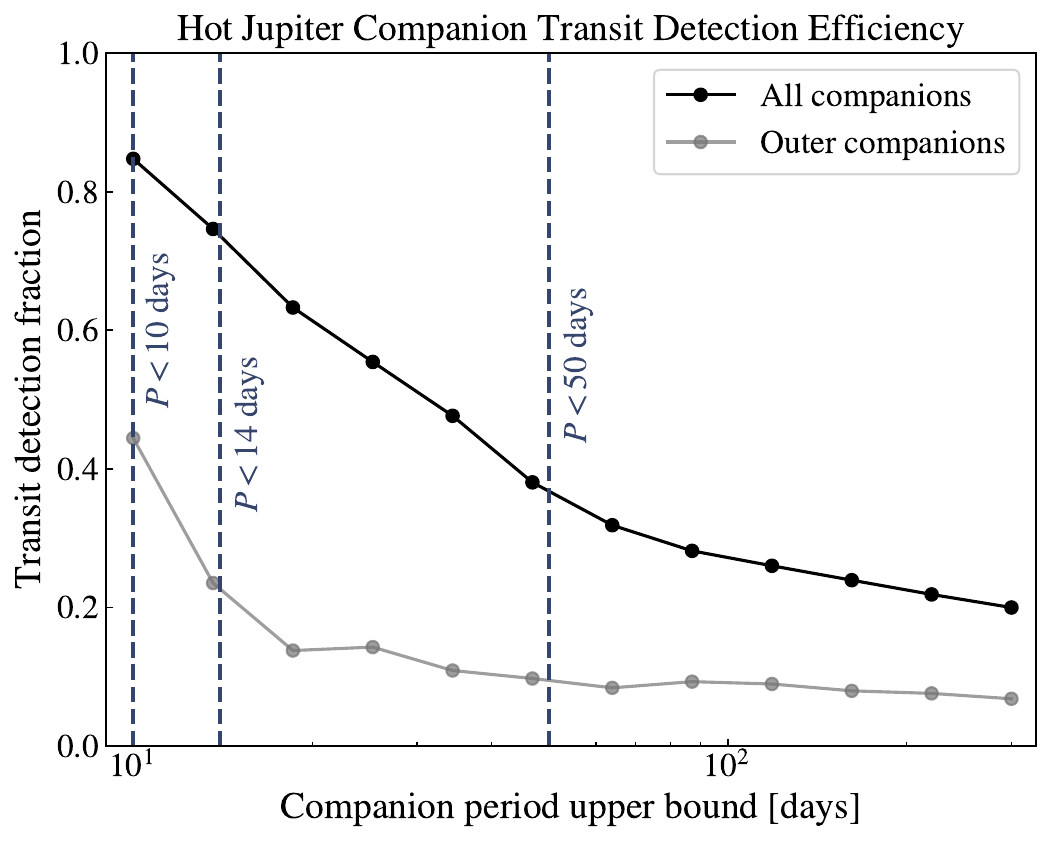}
    \caption{Transit detection fraction for our hot Jupiters' companions as a function of the upper bound imposed on the companions' orbital period range (black line), shown also for outer companions only (gray line). Period cutoffs adopted in observational transit-based studies of hot Jupiter companionship are shown as blue vertical dashed lines: $P<10\,\mathrm{days}$ \citep{Sha2026}, $P<14\,\mathrm{days}$ \citep{Hord_2021}, and $P<50\,\mathrm{days}$ \citep{Huang_2016}. The overall transit detection efficiency of hot Jupiters' companions declines as the companion pool extends to longer orbital periods; this drop is particularly stark from $\approx10$--$20\,\mathrm{days}$. While our hot Jupiters' inner companions are largely detectable, most of their outer companions would be hidden to transit surveys.}
    \label{fig:HJcomp_transitfracs}
\end{figure}
The nearby companion rates reported by previous transit-based hot Jupiter companionship studies are each measured over different orbital period regimes which reflect only a subset of our companions' final periods (see Figure \ref{fig:Pfgiant_Pfcomp}). \cite{Huang_2016} limited their analysis of the \textit{Kepler} data to $P<50\,\mathrm{days}$, recovering a companion rate of just $\approx1\%$. Constrained by the shorter baseline of the TESS sectors, \cite{Hord_2021} and \cite{Sha2026} further restricted their companion searches to $P\leq14\,\mathrm{days}$ and $P<10\,\mathrm{days}$, respectively, yielding companion rates of $7.3^{+15.2}_{-7.3}\%$ and $7.6^{+5.5}_{-3.8}\%$, respectively. We plot $f(p_\mathrm{T}>0.5)$ for our hot Jupiters' companions as a function of orbital period baseline (i.e., the imposed upper bounds on the search for companions) out to $300\,\mathrm{days}$ in Figure \ref{fig:HJcomp_transitfracs}, highlighting the bounds probed by literature estimates. While we caution against literal interpretations of our simplistic, representative transit detection fraction values, we note their sharp decline from nearly 90\% out to $P<10\,\mathrm{days}$ (dominated by the $92\%$ of their inner companions that transit) to just $\approx20\%$ out to $P<300\,\mathrm{days}$. In Figure \ref{fig:HJcomp_transitfracs}, we also separately plot $f(p_\mathrm{T}>0.5)$ for hot Jupiters' outer companions, which exhibits systematically lower values compared to the bulk companion population and a sharper decline at shorter period bounds (e.g., from $44\%$ at $P<10\,\mathrm{days}$ to $21\%$ at $P\leq14\,\mathrm{days}$). This indicates that a large fraction of our hot Jupiters' outer companions may be missed by transit surveys, even those limited to very short baselines. 

\subsubsection{Transit Timing Variations}
\label{subsubsec:TTVs}
Nearby companions may cause their giant planet to undergo TTVs which may be detectable regardless of whether the companions are co-transiting.
To characterize companion detectability via TTVs, we evaluate the amplitudes of the TTV signals exhibited by the giant planets in our simulations. We integrate each system for an additional 500 orbits with \texttt{rebound} and explicitly compute the ``observed'' transit timings of our giant planets (with a precision of $\leq10$ seconds) assuming the observer is placed in the positive $x$-direction. We then take the difference between these observed values and their expected Keplerian values and calculate their root-mean-square to obtain the TTV amplitude $V_\mathrm{TTV}$.

The middle panel of Figure \ref{fig:detectability_CDFs} features the resulting $V_\mathrm{TTV}$ CDFs for hot and warm Jupiters, which our AD test shows are also statistically distinct ($p\leq10^{-4}$). We take $V_\mathrm{TTV}\geq30\,\mathrm{min}$, which corresponds to the median TTV amplitude of the \emph{Kepler} catalog constructed by \cite{Holczer2016}, as the threshold to detect a nearby companion (though we do not distinguish which companion(s) would be detected). The resulting TTV detection fraction $f(V_\mathrm{TTV}>30\,\mathrm{min})$ is $9.6\pm3.0\%$ for hot Jupiters and $33.1\pm2.8\%$ for warm Jupiters, reflecting another significant ($|z\mathrm{-score}|=5.6$) observational dichotomy disfavoring detection for hot Jupiter companions.

These results are consistent with the work of \cite{Wu_2023}, which utilized the full 17-quarter \textit{Kepler} dataset to search for significant TTV signals around short-period giants, deriving nearby companion rates (lower limits) for hot Jupiters and warm Jupiters of $\geq12\pm6\%$ and $\geq70\pm16\%$, respectively. However, we note that our $f(V_\mathrm{TTV}>30\,\mathrm{min})$ values represent only a limited view into our companions' true TTV detectability and also include contributions from all companions in our systems, while the calculations of \cite{Wu_2023} were restricted to a companion-giant period ratio range of $1.5<P_\mathrm{out}/P_\mathrm{in}<4$ (see Figure \ref{fig:Pfgiant_Pfcomp} for reference). Regardless of these differences, we find that $V_\mathrm{TTV}$ scales roughly exponentially with increasing orbital period of the giant, as found in \cite{Wu_2023} (see also the TTV-period relation explored by \citealt{Mazeh2013}). These findings underscore the inherent bias of the TTV method towards detecting companions around longer-period planets\footnote{We note, however, that this TTV amplitude bias may be balanced somewhat by the inherently longer TTV periods exhibited by longer-period planets, which in turn require longer observational baselines to detect a signal at a fixed signal-to-noise ratio.}, generally supporting the notion that companion rates based on the TTVs of short-period planets, especially hot Jupiters, represent only lower limits.

\subsubsection{Radial Velocities}
\label{subsubsec:RVs}
\begin{figure*}
    \centering
    \includegraphics[width=0.9\textwidth,scale=1.0]{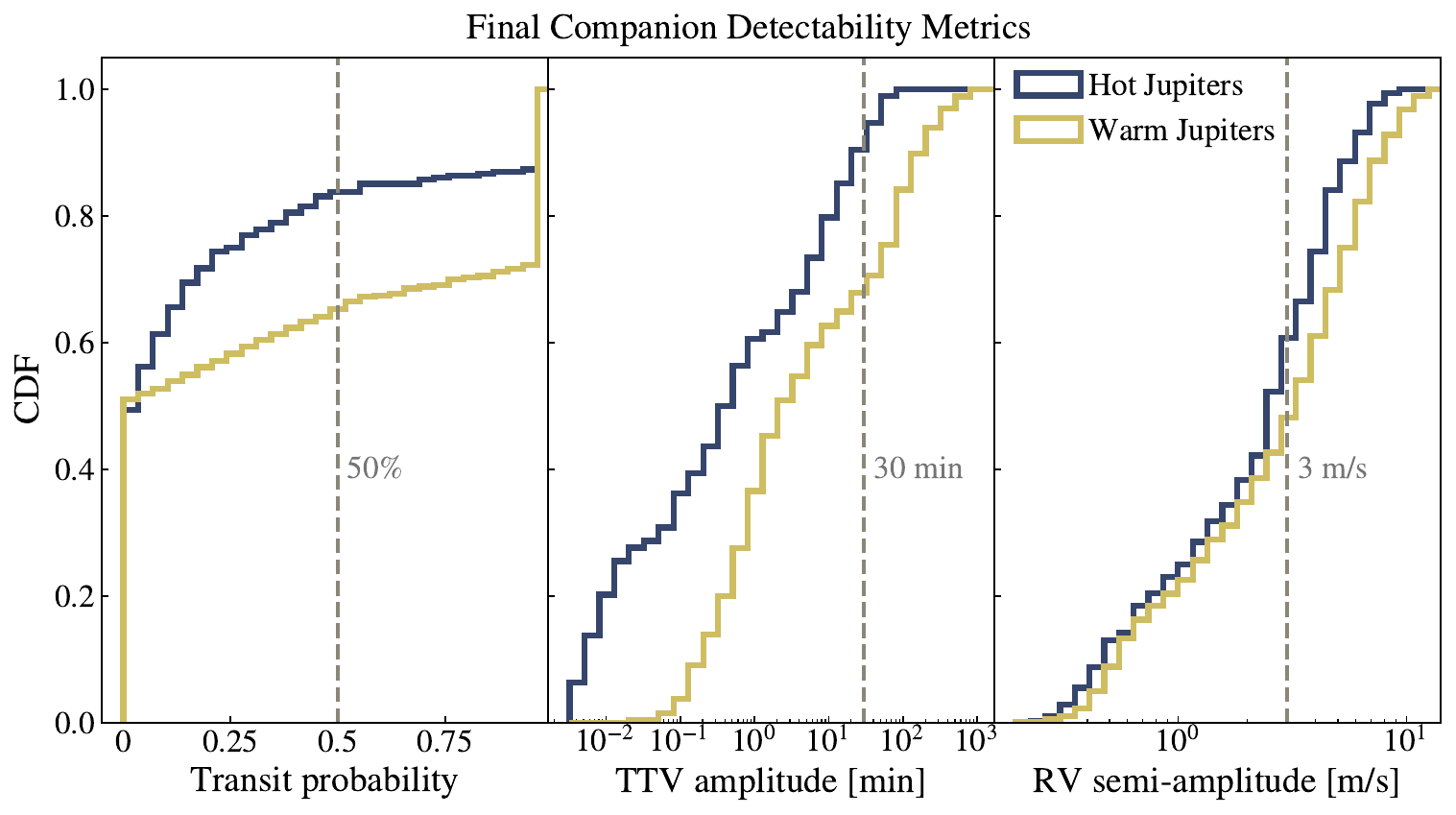}
    \caption{CDFs for the transit probabilities (left), total TTV amplitudes (middle), and RV semi-amplitudes (right) of all remaining companions to hot Jupiters (blue) and warm Jupiters (yellow). In each panel, the vertical dashed lines correspond to the respective thresholds for ``detection'' that we adopt in our comparative analysis: $p_\mathrm{T}>0.5$ for transits, $V_\mathrm{TTV}>30\,\mathrm{min}$ for TTVs, and $K_\mathrm{RV}>3\,\mathrm{m/s}$ for RVs.}
    \label{fig:detectability_CDFs}
\end{figure*}

For our RV analysis, we calculate the analytic RV semi-amplitude $K_\mathrm{RV}$, i.e., reflex motion, induced by each planet on the star, and plot their CDFs for hot Jupiters and warm Jupiters in the right-hand panel of Figure \ref{fig:detectability_CDFs}. Again, we find that these distributions are unique ($p\leq10^{-4}$) and that the companions of warm Jupiters produce larger signals than those of hot Jupiters, though this discrepancy is modest in comparison to what is found for the transit and TTV metrics. Adopting a detection threshold of $3\,\mathrm{m/s}$, which corresponds roughly to the true precision floor achieved by past RV companion searches for short-period giants (see, e.g., the Friends of Hot Jupiters survey; \citealt{Knutson2014} and the TESS Grand Unified Hot Jupiter Survey \citealt{Yee2025}), we find that $43.8\pm2.8\%$ of hot Jupiters host detectable companion(s) compared to $55.5\pm1.7\%$ of warm Jupiters (corresponding to $|z\mathrm{-score}|=3.6$). Therefore, while not completely without bias, we conclude that the RV method may be best to conduct a fair, complete search for nearby companions around hot and warm Jupiters.

\section{Fiducial Simulation Dynamics}
\label{sec:dynamics}

\subsection{A Dichotomy in Instability Outcomes}
\label{subsec:HJ_WJinstab_outcomes}
\begin{figure*}[htb]
    \centering
    \includegraphics[width=1.0\textwidth,scale=1.0]{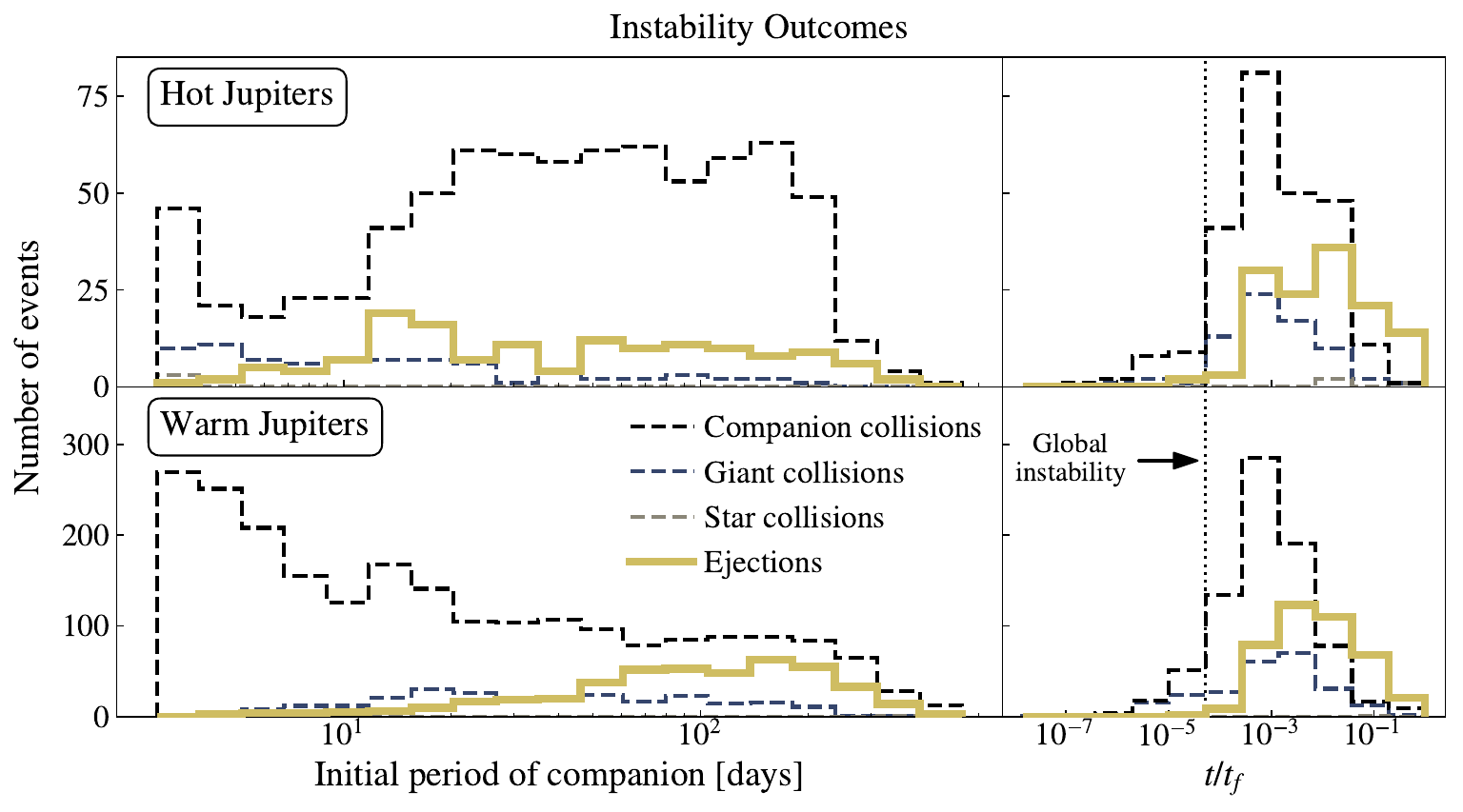}
    \caption{Left: Binned instability outcome counts for our companions as a function of their initial orbital period, plotted separately for hot Jupiter systems (top) and warm Jupiter systems (bottom). Right: Binned fractions of companion instability outcomes as a function of integration time $t$ normalized by the final time $t_f$, plotted for both hot Jupiters (top) and warm Jupiters (bottom). In all panels, companion-companion collisions (black dashed line), companion-giant collisions (blue dashed line), companion-star collisions (gray dashed line), and companion ejections (yellow solid line) are displayed. Instability outcomes vary according to the location of the giant planet in our multi-super-Earth systems, especially for our warm Jupiter systems. Driven by nominal propagation of high eccentricities along the exterior planet chain, the companions of hot Jupiters experience ejection more globally than warm Jupiters' companions, resulting in final period ratio and mutual inclination distributions with larger dispersions.}
    \label{fig:instab_outcomes}
\end{figure*}

All 360 of our fiducial 12-planet systems experience a dynamical instability that ultimately reduces the companion multiplicity.
To characterize their dynamical histories, we track all multiplicity-changing events throughout the simulations. We identify four unique outcomes, each involving our small-planet companions: 1) collision with the host star, 2) collision with the giant, 3) collision with another companion, or 4) ejection from the system. While collisions generally dominate the inner regions and ejections become increasingly common in the outer regions where Safronov numbers \citep{Safronov1972epcf.book.....S} are higher (e.g., \citealt{Bhaskar2025}), we find that the overall picture of dynamical evolution is more complex as the position of the giant planet also bears strong influence over these outcomes. We describe this evolution below, considering separately the case of hot Jupiter and warm Jupiters, whose instability outcomes are displayed in Figure \ref{fig:instab_outcomes}. In brief, we find that the comparatively short orbital periods of hot Jupiters relative to warm Jupiters affords them greater influence over their systems' global evolution. Critically, in hot Jupiter systems, a larger fraction of close-in planets achieve high-eccentricity orbits and are subsequently ejected, resulting in a final companion population that is more distant and mutually inclined, as summarized in Section \ref{subsec:isolation_10days}.

\textbf{Hot Jupiter Systems.} The hot Jupiters in our simulations initially reside at the innermost four positions $j=1$--$4$ and thus host long consecutive chains of 8--11 exterior small-planet companions. The giant quickly excites the eccentricity of its immediate neighbors to large values $\gg e_\mathrm{c}$ (where $e_\mathrm{c}$ is the orbit crossing eccentricity of the system, see Equation \ref{eq:ecross}). Due to its comparatively large mass, the giant's eccentricity remains near zero, so these neighbors must to reach $e \approx 2e_\mathrm{c}$ to collide with it, but only $e\approx e_\mathrm{c}$ to collide with the next-closest small planet, provided that planet also reaches $e\approx e_\mathrm{c}$. For our hot Jupiter systems, $P_\mathrm{giant}$ is short relative to the orbital periods of most companions, thereby setting the characteristic dynamical timescale of the system. As a result, the hot Jupiter dominates its system's global evolution, ensuring that eccentricity excitation is communicated effectively along the chain of outer companions. Its immediate neighbor, therefore, favors collision with the next-closest companion over collision with the giant $\sim$2:1 (a smaller fraction instead undergo ejection).   

What follows next, occurring around $\sim10^3\,\mathrm{yr}$ (i.e., at a fractional simulation time of $t/t_f\sim6\cdot10^{-5}$), is a phase of global instability (see right-hand panels of Figure \ref{fig:instab_outcomes}). Due to its relatively short secular cycle, the hot Jupiter is particularly effective at dynamically stirring its outer companions to high eccentricities $\gg e_\mathrm{c}$ \cite[e.g., Equation 82 of ][]{Pu2021}, resulting in a chaotic cascade of collisions and planet ejections along the chain. As seen in the left-hand panel at the top of Figure \ref{fig:instab_outcomes}, eccentricity excitation and transfer is sufficient to allow companions initially very distant from their hot Jupiter (e.g., $P\gtrsim100\,\mathrm{days}$) to collide with it, and those initially on very close-in orbits to the star ($P\lesssim 10\,\mathrm{days}$) to be ejected.

Although the eccentricities of post-merger planets are damped by the collisional process (e.g., \citealt{Matsumoto2017}), the presence of many other nearby, eccentric planets along the outer chain and persistent dynamical forcing from the inner hot Jupiter help to induce further excitation and instabilities at later times (beyond $t/t_f\gtrsim10^{-2}$; see top right panel of Figure \ref{fig:instab_outcomes}). The ejection of planets along the inner edge of the chain is particularly effective at inducing late instabilities among longer-period companions whose orbits are crossed in the process, as well as clearing out the region near the hot Jupiter. Altogether, these factors help to clear out the vicinity of the hot Jupiter, skewing the final companion-giant period ratio distribution to larger values.

\begin{figure}[htb]
    \centering
    \includegraphics[width=0.45\textwidth,scale=0.75]{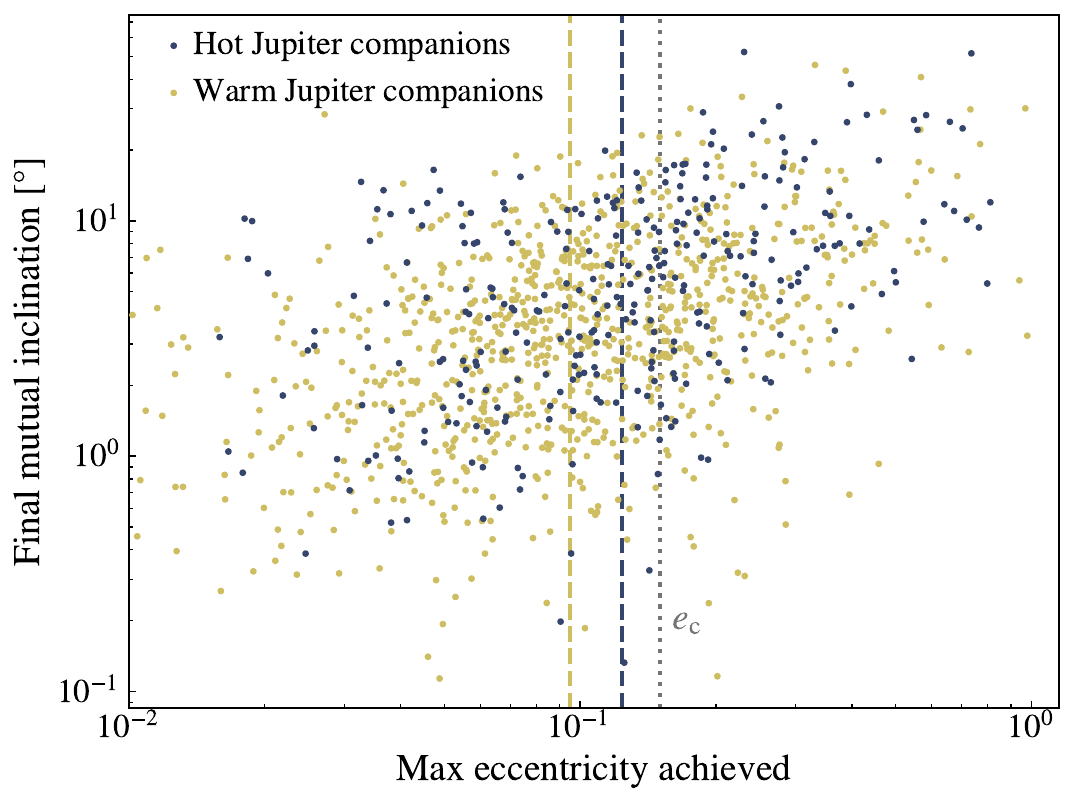}
    \caption{The final mutual inclinations of each surviving giant companion as a function of the maximum eccentricity reached by that companion (after any mergers), plotted separately for the companions of hot Jupiters (blue dots) and warm Jupiters (yellow dots). Median maximum eccentricities for hot Jupiters' and warm Jupiters' companions are indicated via blue and yellow vertical dashed lines, respectively, and the typical initial orbit-crossing eccentricity is represented as the gray vertical dotted line. The strong visible correlation demonstrates that larger eccentricities, which are more often achieved by hot Jupiters' companions, lead to inflated mutual inclinations.}
    \label{fig:Maxecc_Imuts}
\end{figure}

For the case of scattering between many $\sim$equal mass planets, higher eccentricities are more readily `converted' into inclination (see, e.g., the analogous numerical experiment of \citealt{Ida1992}). Therefore, the companions exterior to our hot Jupiters are also driven to orbits with high mutual inclinations--- we demonstrate this eccentricity-inclination coupling in Figure \ref{fig:Maxecc_Imuts}. The final breakdown of instability outcomes for our hot Jupiter systems are: 0.4\% companion-star collisions, 11.0\% companion-giant collisions, 70.8\% companion-companion collisions, and 17.8\% companion ejections.

\textbf{Warm Jupiter Systems.} While still a major dynamical influence in their systems, warm Jupiters have a secular cycle that is longer than hot Jupiters. Specifically, since their orbital periods are longer relative to their companion population, $P_\mathrm{giant}$ represents a less important dynamical timescale, now competing with the timescale set by the period of the innermost planet $P_\mathrm{inner}$ \citep{Funk2010}. The dynamical evolution of these systems thus proceeds somewhat differently. First, a higher fraction of their nearest neighbors are ejected due to the increased relative efficiency of ejections at longer periods \citep{LI2020, Pu2021}. The global instability phase that ensues after is again marked by an array of collisions and ejections, though the occurrence of these events is now a strong function of the companions' initial period (see bottom left panel of Figure \ref{fig:instab_outcomes}). This is because the longer orbital periods of warm Jupiters diminish their ability to drive eccentricity excitation inward, so when $P_\mathrm{inner}\ll P_\mathrm{giant}$ (especially true for warm Jupiters at $j\approx9$--$12$), the evolution of the inner system is dominated by innermost companion, not the giant planet. With minimal dynamical forcing from the warm Jupiter, it is difficult for these interior planets to achieve $e\gg e_\mathrm{c}$, so companion-companion collisions dominate the inner regions during this phase. Consequently, since ejections and giant collisions are largely limited to companions from the outer regions, the inner companion chains remain relatively compact with damped eccentricities and inclinations (Figure \ref{fig:Maxecc_Imuts}).

Driven by the relatively quiescent evolution for the inner chains, the final companion period ratio and mutual inclination distributions for warm Jupiter systems exhibit less scatter and are skewed towards lower values compared to hot Jupiter systems. The final breakdown of instability outcomes for our warm Jupiter systems is: $<0.1\%$ companion-star collisions, 12.8\% companion-giant collisions, 66.8\% companion-companion collisions, and 20.4\% companion ejections. 

\subsection{Hot Jupiters' Inner Companions Become Dynamically Isolated}
This story of dynamical evolution can naturally be applied to explain the ubiquitous population of inclined, outer companions to our hot Jupiters (Section \ref{subsubsec:outer_closein_HJcomps}), but what about the inner, coplanar companions found around our longer-period ($P\approx4.5$--$10\,\mathrm{days}$) hot Jupiters (Section \ref{subsubsec:sweet_spot})? Owing to the initial positions of these hot Jupiters within their systems ($j=2$--$4$), these giants begin post-disk evolution with one, two, or three inner companions. In the case of multiple interior planets, these short chains will undergo dynamical instabilities, quickly reducing themselves to one planet largely through some combination of collisions (see top left panel of Figure \ref{fig:instab_outcomes}). Since the giant acts to dynamically isolate planets on opposing sides of it, there is no mechanism to significantly excite the eccentricity (or mutual inclination) of the remaining inner companion, and it will effectively become part of a two-body system with the giant. The resulting spacing is of order $K' \sim 2K \gg 2\sqrt{3}$, so the pair will be Hill stable indefinitely and the hot Jupiter will retain this nearby inner companion throughout its system's evolution (see also \citealt{Bhaskar2024}). Similarly, we find that for the single-companion case, the inner planet remains stable 27\% of the time, with a mean initial Hill spacing of $K=5.5$.

\subsection{Final Stability}
Our simulation cut-off criteria of $10^9$ giant orbits or $20 \,\mathrm{Myr}$ were selected to ensure that mature and comparable dynamical ages were reached for both hot and warm Jupiter systems while mitigating computational cost. Adopting a fixed integration time in terms of $P_\mathrm{giant}$, which varies by $
\sim$two orders of magnitude, would ignore the increasing influence of $P_\mathrm{inner}$ (which is constant across simulations) as a dynamical timescale for longer-period giant systems. Similarly, cutting off the simulations at a fixed integration time would also correspond to a fixed multiple of $P_\mathrm{inner}$ and neglect the dynamical influence of our giant planets, which is particularly dominant for those at the shortest orbital periods (Section \ref{subsec:HJ_WJinstab_outcomes}). Hence, the dual cut-off criteria help to optimize for these competing timescales.

The general alignment between the fractional global instability times of our hot Jupiter and warm Jupiter systems (at $t/t_f\sim6\cdot10^{-5}$; see right-hand panels of Figure \ref{fig:instab_outcomes}) suggests they have achieved comparable dynamical ages. To better investigate the final stability of our systems, we use the predictive capabilities of the \texttt{SPOCK} machine-learning tool (Stability of Planetary Orbital Configurations Klassifier; \citealt{TamayoSPOCK2020}). Specifically, utilizing the Deep Regressor model \citep{Cranmer2021}, we calculate a mean instability probability of 40.8\% for our hot Jupiter systems and 39.2\% for our warm Jupiter systems (predicted over the next $10^9P_\mathrm{inner}$), verifying that both populations of giants have reached similar dynamical ages. Although these absolute probabilities may not be very accurate for our final systems (whose architectures deviate significantly from the well-ordered, equal-mass super-Earth like systems used to train \texttt{SPOCK}; \citealt{Tamayo2016,Tamayo2020}), these results indicate that a substantial fraction of our systems may still experience further instabilities. This is supported by the non-negligible frequency of ejection events observed in our simulations beyond $t/t_f\gtrsim10^{-1}$ (Figure \ref{fig:instab_outcomes}). Since such late-time instabilities are most likely to result in planet ejections, we expect that extending these simulations may further exaggerate the hot Jupiter-warm Jupiter period ratio and mutual inclination dichotomy.

\section{Robustness Experiments}
\label{sec:nonfiducial_results}
We present non-fiducial simulations with alternate setups exploring the robustness of our results to the adopted planet-planet spacing and the inclusion of planetary tidal forces in Sections \ref{subsec:nonfiducial_spacing} and \ref{subsec:tides}, respectively. 

\subsection{Dependence on Initial Spacing}
\label{subsec:nonfiducial_spacing}
Since the primordial compactness of our simulated systems, set by our adopted $K$ value, is expected to play a significant role in shaping the systems' evolution, we run two additional suites of 360-system simulations with modified planet spacings: a more compact suite with $K=8$ (see e.g., \citealt{Kokubo1995,Kokubo1998, Kokubo2000}) and a less compact suite with $K=12$. All other initial conditions were determined following the same procedures implemented for our fiducial suite, described in Section \ref{sec:simulation_setup}. In order to mitigate computational cost, the maximum integration time for these simulations is capped at $10^9P_\mathrm{giant}$ or $10\,\mathrm{Myr}$ (compared to  $20\,\mathrm{Myr}$ for our fiducial simulations).  

We present the main results of these non-fiducial simulations in Appendix \ref{app:nonfid_results}. Ultimately, we find that our our main conclusions hold in both suites: hot Jupiters' companions end up on orbits with statistically larger companion-giant period ratios and mutual inclinations than warm Jupiters' companions (Section \ref{subsec:isolation_10days}), and consist of both an inner, coplanar and outer, inclined subpopulation (Section \ref{subsec:architectures}). Further, for each of the transit, TTV, and RV detectability metrics investigated here, we again find that a lower fraction of hot Jupiter companions are detectable compared to warm Jupiter companions (Section \ref{subsec:obs_dichotomy}). Therefore, the production of preferentially isolated hot Jupiters appears to be robust to the precise initial spacing adopted. We note, however, that the significance of this dichotomy in companionship properties appears to depend somewhat on $K$, potentially strengthening for wider initial spacings. We discuss this phenomenon further in Appendix \ref{app:nonfid_results}.

\subsection{The Influence of Planetary Tidal Forces}
\label{subsec:tides}
For simplicity and to isolate the role of dynamical interactions in producing the companionship properties of our hot and warm Jupiters, we have excluded additional forces in our fiducial simulations. However, we note that tidal forces raised on the planets by the central star may play a significant role in sculpting the final architectures of our systems, potentially rapidly de-coupling highly eccentric small planets from their systems during phases of instability \citep{Marzari2019} or otherwise damping companion eccentricities and inclinations. Over longer timescales ($\sim\mathrm{Gyr}$), tides may also drive long-term inward migration for our innermost ($P\lesssim3 \, \mathrm{days}$) hot Jupiters (e.g., \citealt{Rasio_1996, Levrard_2009}) or even their companions \citep{Pu2019}.

To investigate this effect, we include tidal forces using the \texttt{tides\_constant\_time\_lag} model \citep{Baronett2022} implemented in \texttt{reboundx} \citep{Tamayo2020} with otherwise identical initial conditions to our fiducial suite (i.e., with $K=10$). We adopt Love numbers $k_2$ of $0.55$ and $0.15$ for our giants and small planets, respectively (see \citealt{BeckerBatygin2013}). The constant time lag parameter, which is updated regularly for each planet throughout the simulation, is calculated as $\tau=1/2nQ$ \citep{Lu2023}, where $n$ is the orbital frequency and $Q$ is the tidal quality factor, set to $10^5$ for our giants and $10^4$ for their companions \citep{Millholland2019}. As is the case for our $K=8$ and $K=12$ non-fiducial simulations, we cut off these simulations at the lesser of $10^9P_\mathrm{giant}$ or $10\,\mathrm{Myr}$.

In Appendix \ref{app:nonfid_results}, we describe the key companionship properties of our final hot Jupiter and warm Jupiter systems with this tidal prescription implemented. As seen for our $K=8$ and $K=12$ non-fiducial simulations, our key results remain unchanged--- we again find that hot Jupiters' companions end up with larger period ratios and mutual inclinations, and lower detection fractions via the transit, TTV, and RV methods. Our hot Jupiters also end up with nearby companions on both coplanar, interior orbits and mutually-inclined, exterior orbits. Further, we find that the companion statistics compiled for this suite are each in good agreement with the results from our fiducial $K=10$ suite (see Table \ref{tab:results_summary}), with the notable exception of mutual inclinations, which are systematically lower for the hot Jupiter systems in our tidal suite. We conclude this is caused by tidal damping and show that it does not significantly alter the transit detection fractions for hot Jupiters' companions. Tides may continue to play a minor role in our systems' longer-term evolution, particularly through our companion planets (e.g., $\sim30\%$ have dissipation timescales $\leq1\,\mathrm{Gyr}$).

\section{Compatibility With Quiescent Formation Mechanisms}
\label{sec:compatability}
This study seeks to explore the isolation dichotomy produced between hot Jupiters and warm Jupiters from N-body dynamics assuming a unified, quiescent formation scenario.  
As such, we do not specify the mechanism by which our giant planets can emerge from the disk embedded in the high-multiplicity, compact super-Earth systems that we simulate here, but we briefly discuss the compatibility of our initial conditions with both in situ formation and disk-driven migration below.

\textbf{In Situ Formation.} Considering first the in situ formation scenario, it has been demonstrated that the critical mass required for a rocky core to undergo runaway gas accretion \citep{Pollack1996} is only weakly dependent on nebular conditions \citep{STEVENSON1982, Rafikov2006}, meaning that giant planet formation may be viable across a broad range of orbital separations \citep{Lee2014}. Since the runaway accretion process is expected to be relatively rapid, even close in to the star ($<10^5\,\mathrm{yr}$; e.g., \citealt{LeeChiang2015,Batygin2016}), in-situ formation for giants hinges largely on whether a critical-mass core of $\sim10$--$30\,\mathrm{M}_\oplus$ can be formed before the disk dissipates. The solid accretion timescale of the core and the disk removal timescale depend on many variables, including details of the disk's structure and composition. Provided there is a sufficient reservoir of solid material in the innermost regions of the disk (see \citealt{Chiang_Laughlin2013,Schlichting2014}), there are several promising mechanisms to assemble massive super-Earth cores on timescales $\lesssim1\,\mathrm{Myr}$ \citep{Lambrechts2012, Lambrechts2014, Lenz_2019, Zawadzki2021} --- i.e., well within the typical disk lifetime of $1$--$10\,\mathrm{Myr}$ (\citealt{Haisch2001, Jayawardhana2006, Currie2009}). This is especially true for more massive disks, which not only likely have more solid mass but also generally have longer lifetimes \citep{Bayo2012, Ribas2015, Ben2025}, potentially lowering the critical core mass needed for runaway accretion \citep{LeeChiang2015}.

Regardless, the tight orbital configurations of our giant planets within their multi-planet systems may trigger an instability that halts the runaway accretion process, preventing such an architecture from being realized. While the random enhancements we introduce to the mutual Hill spacing of our giant planets ($\delta K\sim K$) help to account for this, we find that the time for one or more companions to reach $e=e_\mathrm{c}/2$, which we take as the dynamical instability growth time $t_\mathrm{dyn}$, is $\sim10$--$100\,\mathrm{yr}$. Although this is short compared to the runaway accretion timescale $t_\mathrm{acc}\sim 10^4$--$10^5\,\mathrm{yr}$ \citep{Batygin2016}, gaseous disks have an eccentricity-damping effect that stabilizes planetary orbits \citep{Ward1997,Chatterjee2008}, even in the presence of a giant \citep{Li2024}. For our super-Earths, the eccentricity-damping timescale $t_\mathrm{damp}$ is expected to be $\sim100\,\mathrm{yr}$ \citep{Goldberg_2022}, which is comparable to $t_\mathrm{dyn}$. In principle, since this damping efficiency decreases linearly with increasing planet mass \citep{Matsumura2010}, $t_\mathrm{damp}$ should be even shorter for our growing giants, ensuring stability throughout the runaway accretion phase. However, if the giant clears a gap in the disk, this damping effect too will vanish, allowing for rapid eccentricity excitation \citep{Moorhead2008, Matsumura2010}, which may stunt the proto-giant's growth. In any case, if the giant is able to form in situ prior to the onset of instability, it seems inevitable that hot Jupiters become observationally isolated relative to warm Jupiters--- either due to long-term post-disk evolution in high-multiplicity systems, as shown in this work, or perhaps sweeping secular resonances in the limit of lower companion multiplicity, as demonstrated in \cite{Batygin2016}.

\textbf{Disk Migration.} A possible contradiction of the in situ model is that rocky planets embedded in a gas-rich disk are subject to Type I orbital migration \citep{Artymowicz1993, Ward1997,Terquem2007, Ida2008, Ida_2010, McNeil2010}, which can only be avoided in a gas-poor or dissipating disk \citep{Raymond2008,Chiang_Laughlin2013, Raymond2014}, but such an environment lacks the necessary conditions for runaway accretion. Further, chains of multiple super-Earths are expected to undergo convergent migration and subsequently lock into first-order MMRs (e.g., 3:2 or 2:1; \citealt{Lee2002, Terquem2007, Raymond2008, McNeil2010}), which are later broken by mutual interactions after the disk disappears \citep{Kominami2004, Iwasaki2006, Izidoro_2017, Izidoro_2021}. Due to our systems' high multiplicity and short instability timescale, the initial proximity of our super-Earths to MMRs, and hence whether they initially migrated in resonant chains, is not expected to have a significant effect on the strength of the final companionship dichotomy between our hot Jupiters and warm Jupiters (see \citealt{MATSUMOTO2012}). The presence of a migrating giant, however, may have important implications.

Disk-driven migration allows proto-hot and -warm Jupiters to form farther out in the disk, near the peak of the observed giant occurrence rate ($\sim 3\,\mathrm{AU}$; \citealt{Fernandes2019,Fulton2021}), where the conditions for core agglomeration are more favorable \citep{Hayashi1981, Pollack1996, Kokubo2002, Masset_2006}, and the instability growth timescale is longer. Continuous angular momentum exchange with the disk may bring the giant inward to short orbital periods on timescales of $10^5$--$10^6\,\mathrm{yr}$, eventually clearing a gap and coming to a stop (likely near the corotation radius \citealt{LeeChiang2017}) as described by the Type II migration paradigm \citep{Goldreich_Tremaine1979, Lin_Papa1986, Lin1993, Lin1996}. 

While super-Earths born interior to the proto-giant will be able to freely migrate inwards (on timescales of $\sim10^5\,\mathrm{yr}$; e.g., \citealt{Goldberg_2022}), the giant acts to block inward-moving exterior companions or their constituent planetesimal building blocks. Consequently, there is expected to be a pile-up of outer companions near the outer edge of the gap (e.g., at the $\sim$2:1 MMR) which is susceptible to strong mutual gravitational interactions that are likely to break the peas-in-a-pod structure \citep{Ogihara2014,Izidoro2015}. In addition to collisions and ejections, such instabilities may result in `jumpers', i.e., outer super-Earths capable of crossing the giant's orbit to end up in the inner disk. This outcome is expected to be more common for outer systems with higher initial multiplicities (e.g., containing $\geq 5$ super-Earths; \citealt{Izidoro2015}). Therefore, disk migration may cause a preliminary clearing of outer companions and also transport some to interior orbits, especially for the innermost giants with the longest outer chains.
The net effect is that our hot Jupiters (and some close-in warm Jupiters) would begin inherently more isolated, though perhaps with an additional companion on the interior.

\section{Summary and Outlook}
\label{sec:conclusions}
In this work, we ran multiple suites of $N$-body simulations using \texttt{rebound} and \texttt{reboundx} to study the companionship properties of short-period giants embedded in high-multiplicity systems of small  super-Earth-like planets.
We have demonstrated that formation and subsequent long-term dynamical evolution in such compact multi-planet systems will naturally result in a preferential observed isolation for hot Jupiters relative to warm Jupiters. Our findings can be summarized by the following key results:
\begin{enumerate}
    \item The final companion-giant orbital period ratios ($P_\mathrm{out}/P_\mathrm{in}$) and mutual inclinations ($\mathcal{I}_\mathrm{mut}$) of hot Jupiters' companions are significantly larger than those of warm Jupiters. While some of the period ratio dichotomy is inherent to the initial architectures of these systems, the difference in mutual inclinations can be attributed fully to dynamical evolution. 
    \item Compared to warm Jupiters, the companions of hot Jupiters are significantly less detectable in terms of their geometric transit probabilities as well as the amplitudes of the transit timing variation and radial velocity signals they produce. This observational bias is weakest for radial velocity, which also yields the highest overall companion detection efficiency. 
    \item Hot Jupiters retain two distinct populations of nearby companions: inner planets on coplanar ($\mathcal{I}_\mathrm{mut}\lesssim2\degree$) orbits and outer planets on inclined orbits. The former population is specific to longer-period hot Jupiters ($4.5$--$10\,\mathrm{days}$) but has a high transit detection probability while the latter population is ubiquitous but difficult to detect via transits, boasting higher mutual inclinations than assumed in most occurrence rate studies. 
\end{enumerate}
Crucially, key results 1 and 2 show that a significant observational companionship dichotomy between hot Jupiters and warm Jupiters can be achieved through a unified quiescent paradigm for their origins. As such, hot Jupiters' preferential isolation is not unique to high-eccentricity migration, and may be attributed in part to disk migration (or in situ formation). This degeneracy provides important context for companionship studies, especially those based on transit and transit timing variation searches.

The growing sample of hot Jupiters with nearby companions found by transit surveys (e.g., \citealt{Canas_2019,Huang_2020,Hord_2022,Maciejewski2023,Sha_2023,Korth2024,Hord2024,McKee_2025,Quinn2026}) affirms that a non-negligible fraction of hot Jupiters \textit{must} form quiescently. These multi-transiting systems are broadly consistent with our simulated hot Jupiter companion architectures, summarized in key result 3. Our simulations demonstrate that additional nearby outer companions may exist undetected in these systems and in many more which are known to contain only a lone hot Jupiter. Although expensive compared to transit surveys, high-precision ($\lesssim1\,\mathrm{m/s}$), moderate-baseline ($\lesssim50\,\mathrm{days}$) radial velocity campaigns are likely necessary to detect these inclined companions and thus definitively calculate the true rate of nearby companions to hot Jupiters.

The compact initial conditions adopted here represent an idealistic and oversimplified view of a unified quiescent origin story for hot Jupiters and warm Jupiters. While we attempt to keep our setup agnostic to any specific quiescent formation scenario, the details underlying how short-period giants may emerge from tightly-packed, high-multiplicity systems are unclear, and we speculate that the true post-formation architectures of these systems may deviate significantly from our assumptions. One complicating but potentially important factor neglected here is the dynamical influence of a distant outer giant perturber, which RV surveys indicate are common in known hot Jupiter systems \citep{Bryan2016,Zink2023}. The addition of such a companion may be addressed in future work.

Finally, while we do not claim that compact quiescent origins are a dominant outcome for hot Jupiters (see e.g., \citealt{Dawson2018} and references therein), our results generally provide more leeway in interpreting their observed isolation as a signature of either dynamically violent or quiescent histories. Recently, using stellar velocity dispersion as a system age indicator, the statistical analysis of \cite{Schmidt2026} found that high-eccentricity migration accounts for at least $40\%$ but at most $70\%$ of the hot Jupiter population (see also the constraints from \citealt{Jackson2023}). Such a result implies that, while still producing only a minority fraction ($\sim1/3$) of these peculiar giants, quiescent origins may indeed be more important than previously thought. Together with our work, this provides further motivation to search for hidden close-in companions to hot Jupiters. Simultaneously, comparisons of the atmospheric compositions of hot and warm Jupiters, enabled by ongoing campaigns with NASA's James Webb Space Telescope (see, e.g., \citealt{Kirk2024,Lindsey2025}) and dedicated large-scale surveys with the European Space Agency's upcoming \emph{Ariel} mission (e.g., see \citealt{DAoust2025}), provide a promising complementary means to constrain the formation histories of both hot Jupiters and warm Jupiters.

\begin{acknowledgments}
We would like to thank Cristobal Petrovich, Andr{\'e} Izidoro, and Dong-Hong Wu for their insightful discussion related to this work, as well as Armaan Goyal for his expertise regarding statistical methods.

This work was supported in part by the NASA Exoplanets Research Program  NNH23ZDA001N-XRP (Grant No. 80NSSC24K0153), the NASA TESS General Investigator Program, Cycle~7, NNNH23ZDA001N-TESS (Grant No. 80NSSC25K7912), and the Heising-Simons Foundation (Grants \#2023-4050, \#2021-2802, and \#2023-4478). We acknowledge funding support from grant JWST-GO-09025.010-A provided by NASA via the Space Telescope Science Institute under the JWST General Observers Program \#9025. This research was supported in part by Lilly Endowment, Inc., through its support for the Indiana University Pervasive Technology Institute.

\end{acknowledgments}

\software{astropy \citep{astropy:2013,astropy:2018,astropy:2022}, matplotlib \citep{Hunter2007}, numpy \citep{numpycitation}, pandas \citep{mckinney-proc-scipy-2010, reback2020pandas}, 
          rebound \citep{rebound}, 
          reboundx \citep{Baronett2022}, SPOCK \citep{TamayoSPOCK2020} 
          }

\appendix

\section{Appendix A: Energy and Angular Momentum Conservation}\label{app:energy_AM}
\texttt{TRACE} \citep{reboundtrace} is a hybrid integrator that utilizes the symplectic Wisdom-Holman integrator \texttt{WHFast} \citep{Rein_whfast2015} for secular interactions and switches to a higher-order scheme when the minimum mutual Hill separation between any two bodies falls below a certain threshold. To ensure we achieve the best possible accuracy during phases of instability, we adopt a conservative switching threshold of $r_{H,\mathrm{mut}}<7$, and switch to the 15th-order adaptive-step \texttt{IAS15} integrator \citep{reboundias15} rather than the adaptive-step (and adaptive-order) Gragg-Bulirsch-Stoer scheme \citep{BS_integrator}. We retain the default value of $10^{-9}$ for \texttt{IAS15}'s dimensionless precision parameter $\epsilon_b$, which corresponds to a nominal relative energy error after one timestep of $8.9\times10^{-34}$ (see Equation 23 of \citealt{reboundias15}), though in practice the minimum relative error has a floor of $10^{-16}$. We note that this default value for $\epsilon_b$ is selected such that this machine precision is expected to be maintained even after $\sim 10^{10}$ dynamical times (orbits) and with multiple bodies having eccentric orbits \citep{reboundias15}. We set the maximum timestep to $P_\mathrm{giant}/15$ (no minimum timestep is specified). For \texttt{WHFast}, we set the fixed timestep to $1/15$th of the initial innermost period of our systems (consistent with recommendations from \citealt{Rein_whfast2015}). For our fiducial suite, we compute median relative energy and angular momentum errors of $1.5\times10^{-4}$ and $2.8\times10^{-2}$, respectively, and find that the latter is dominated by the removal of ejected planets. The maximum energy error achieved is $3.5\times 10^{-3}$, indicating good overall accuracy. Comparable results are found across all three non-fiducial simulation suites (see Section \ref{sec:nonfiducial_results}), with the median energy and angular momentum errors remaining at $
\sim10^{-4}$ and $
\sim10^{-2}$, respectively.

\section{Appendix B: Non-fiducial Results}
\label{app:nonfid_results}
\begin{figure*}[ht!]
    \centering
    \subfloat[CDFs for our $K=8$ simulations.\label{fig:detectability_CDFs_nonfid_a}]{
        \includegraphics[width=0.6\linewidth]{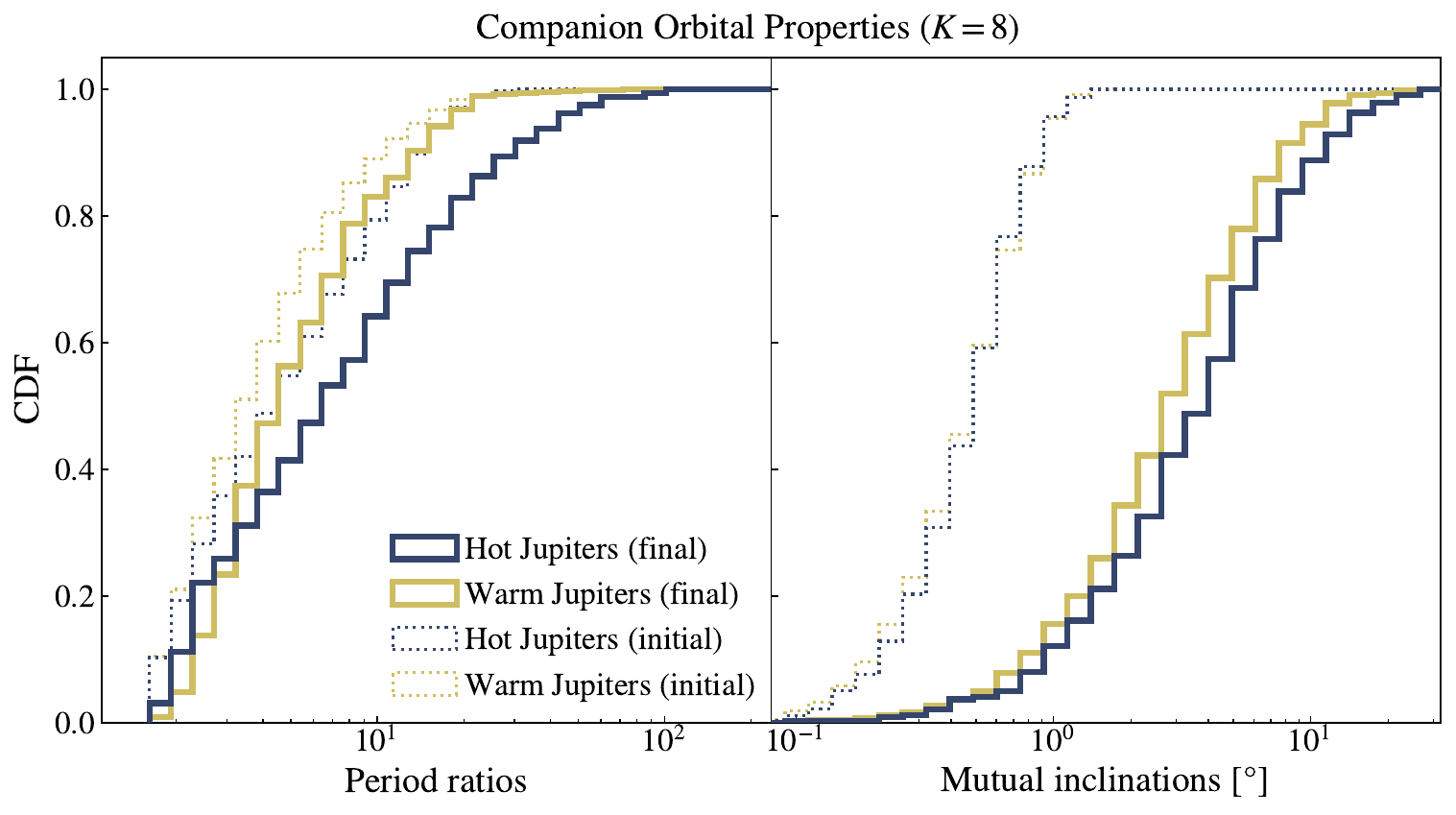}
    } \\ 
    \subfloat[CDFs for our $K=12$ simulations. See panel (a) above for plot legend.\label{fig:PR_RMSImut_CDFs_nonfid_b}]{
        \includegraphics[width=0.6\linewidth]{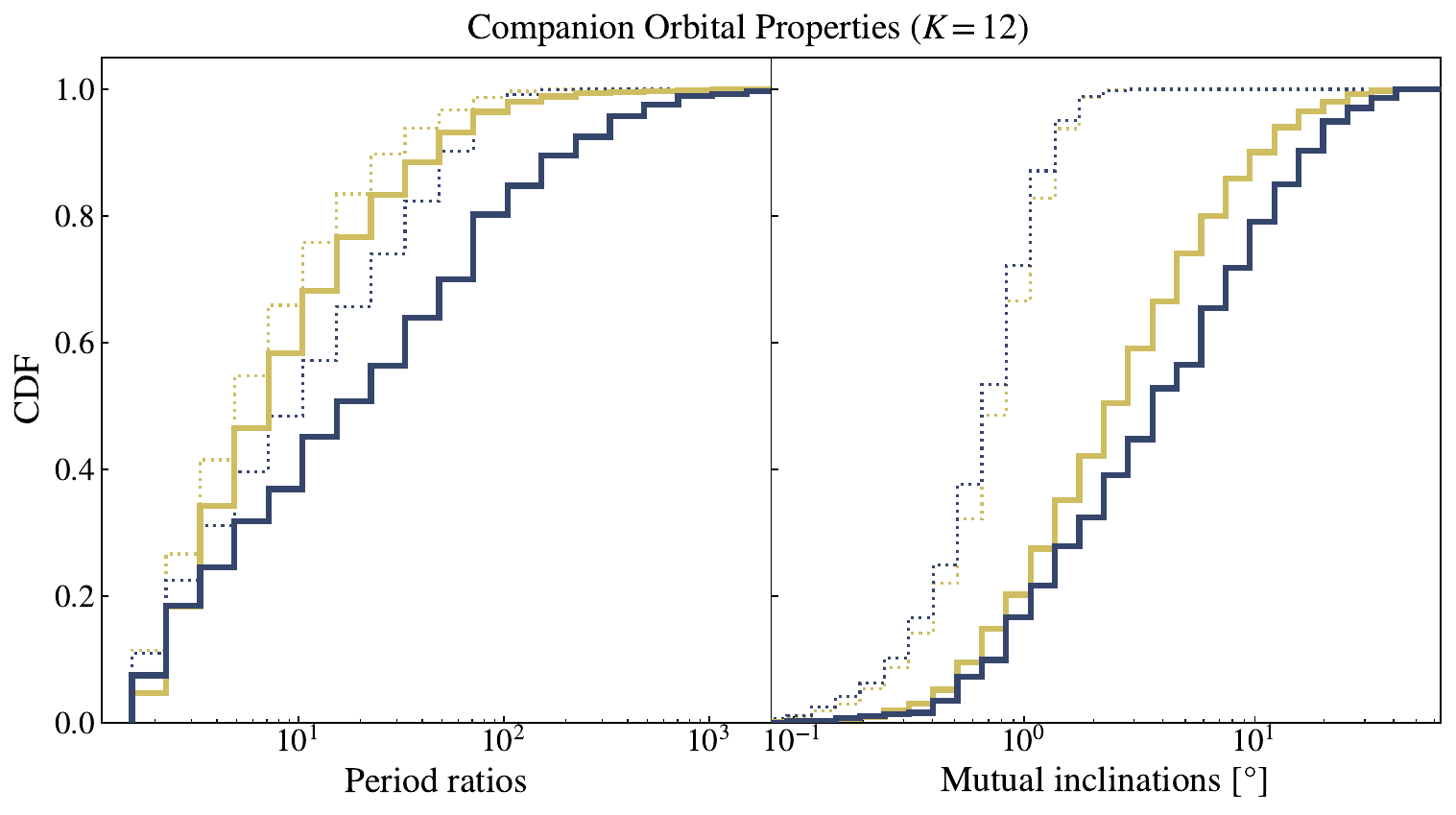}
    } \\ 
    \subfloat[CDFs for our $K=10$ + tides simulations. See panel (a) above for plot legend.\label{fig:PR_RMSImut_CDFs_nonfid_c}]{
        \includegraphics[width=0.6\linewidth]{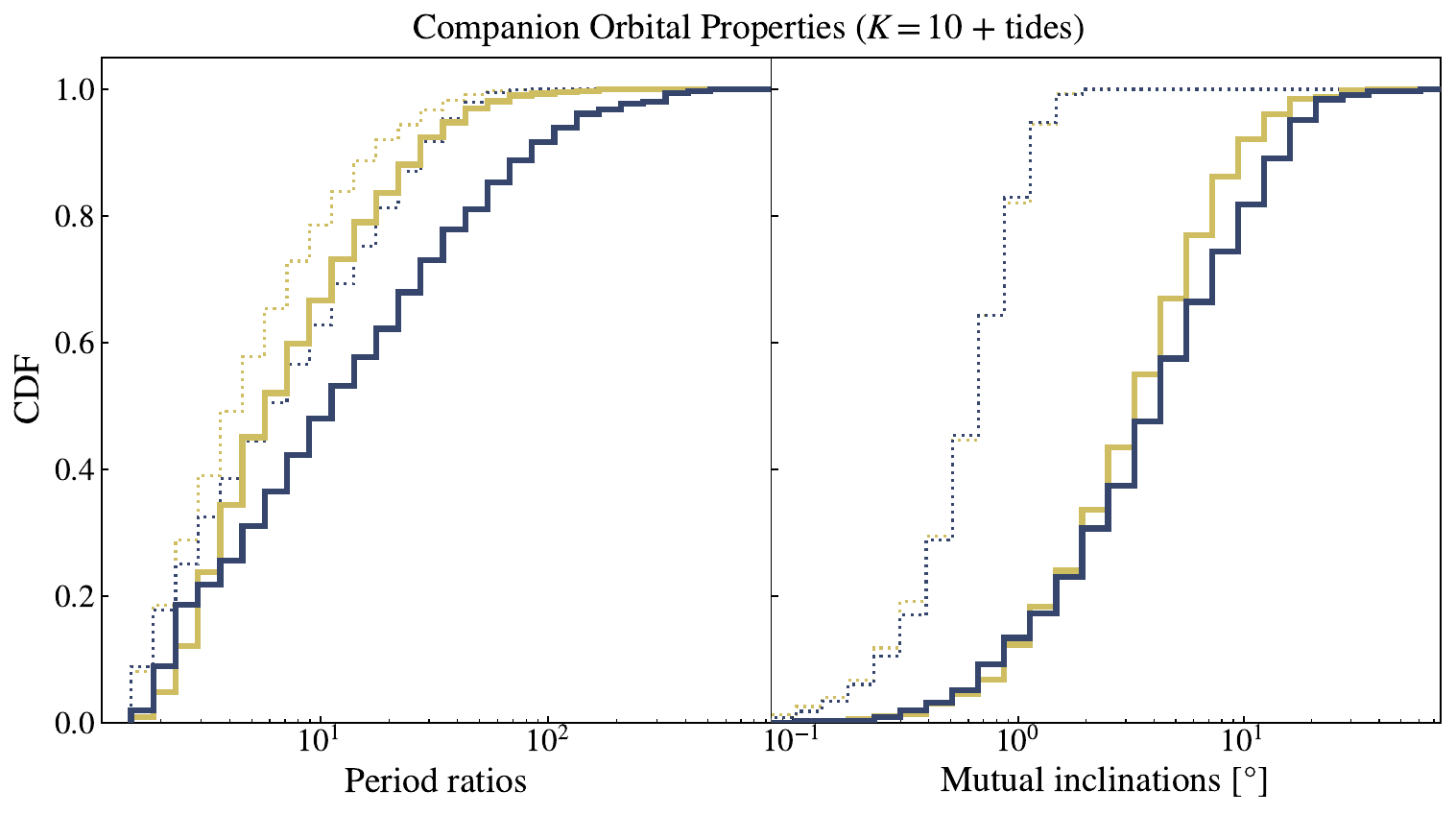}
    }
    \caption{Companion period ratio and mutual inclination CDFs for our non-fiducial simulation suites: (a) $K=8$ results, (b)  $K=12$ results, and (c) $K=10$ + tides results. See caption of Figure \ref{fig:PR_RMSImut_CDFs} for further plot details. In all cases, an exaggerated period ratio dichotomy and a mutual inclination dichotomy form between hot and warm Jupiters.}
    \label{fig:PR_RMSImut_CDFs_nonfid}
\end{figure*}

In addition to our fiducial simulations, we run three additional suites of simulations to test the robustness of our results: two with modified initial spacing parameters ($K=8$ and $K=12$) and one with our original spacing but planetary tidal forces included (see Section \ref{sec:nonfiducial_results}). Repeating our analysis from Section \ref{subsec:isolation_10days} for each suite, we plot the CDFs of our companion period ratios and mutual inclinations in Figure \ref{fig:PR_RMSImut_CDFs_nonfid}, which demonstrates that in each case, hot Jupiters' companions still end up significantly more distant and inclined compared to warm Jupiters' companions. This is corroborated by the large $z\mathrm{-scores}$ for the median values of their respective $P_\mathrm{out}/P_\mathrm{in}$ and $\mathcal{I}_\mathrm{mut}$ distributions, which we verify are statistically distinct via AD tests ($p\leq10^{-3}$ for all tests). We report these median results, including also the distribution means, in Table \ref{tab:nonfidresults_summary}. For each of these non-fiducial suites, we additionally verify that the final companion architectures are qualitatively similar to our fiducial $K=10$ architectures, featuring two distinct populations for hot Jupiter systems: inner, coplanar companions and outer, inclined companions.

\begin{figure*}[ht!]
    \centering
    \subfloat[CDFs for our $K=8$ simulations.\label{fig:detectability_CDFs_nonfid_a}]{
        \includegraphics[width=0.6\linewidth]{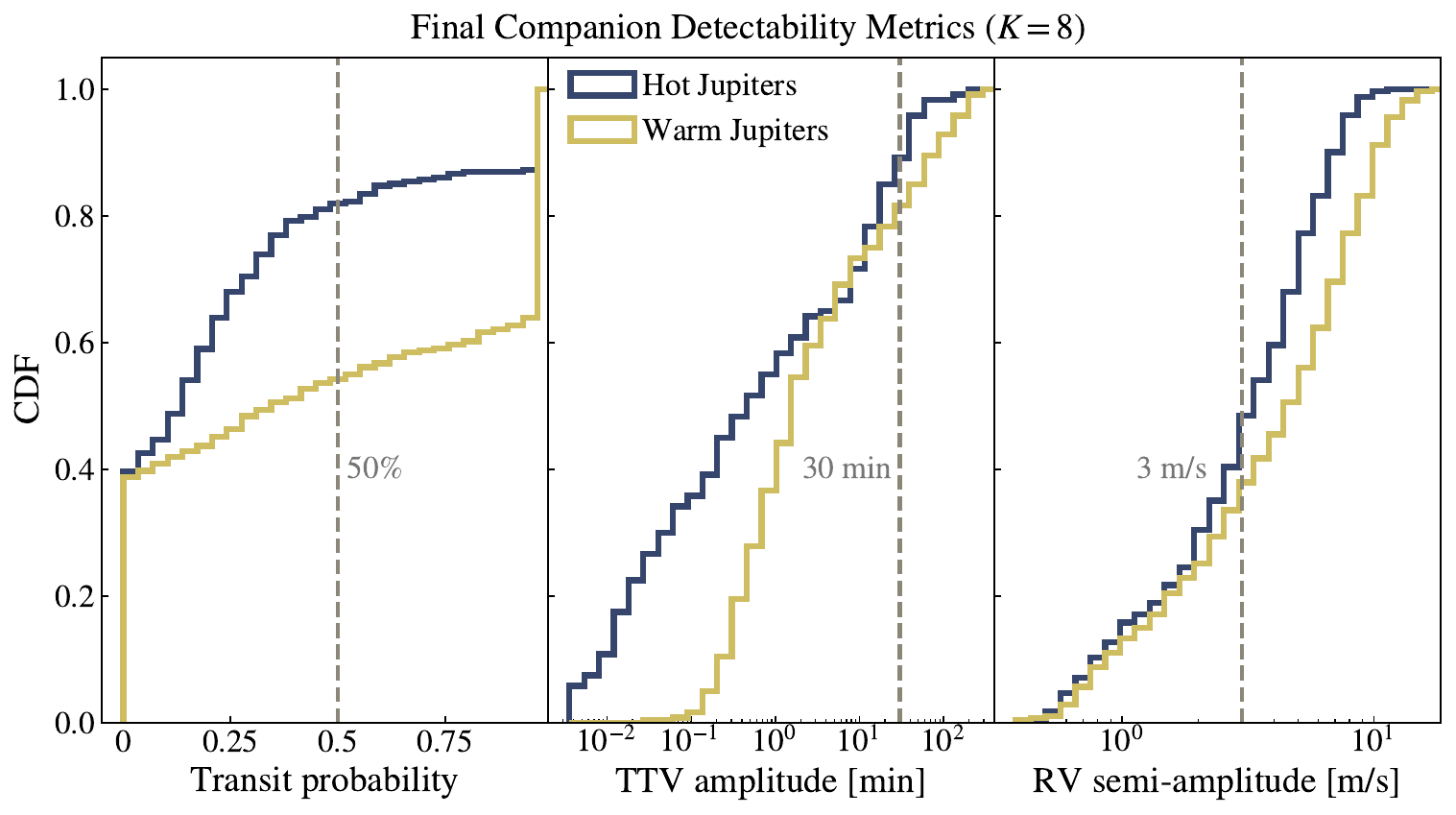}
    } \\ 
    \subfloat[CDFs for our $K=12$ simulations. See panel (a) above for plot legend.\label{fig:detectability_CDFs_nonfid_b}]{
        \includegraphics[width=0.6\linewidth]{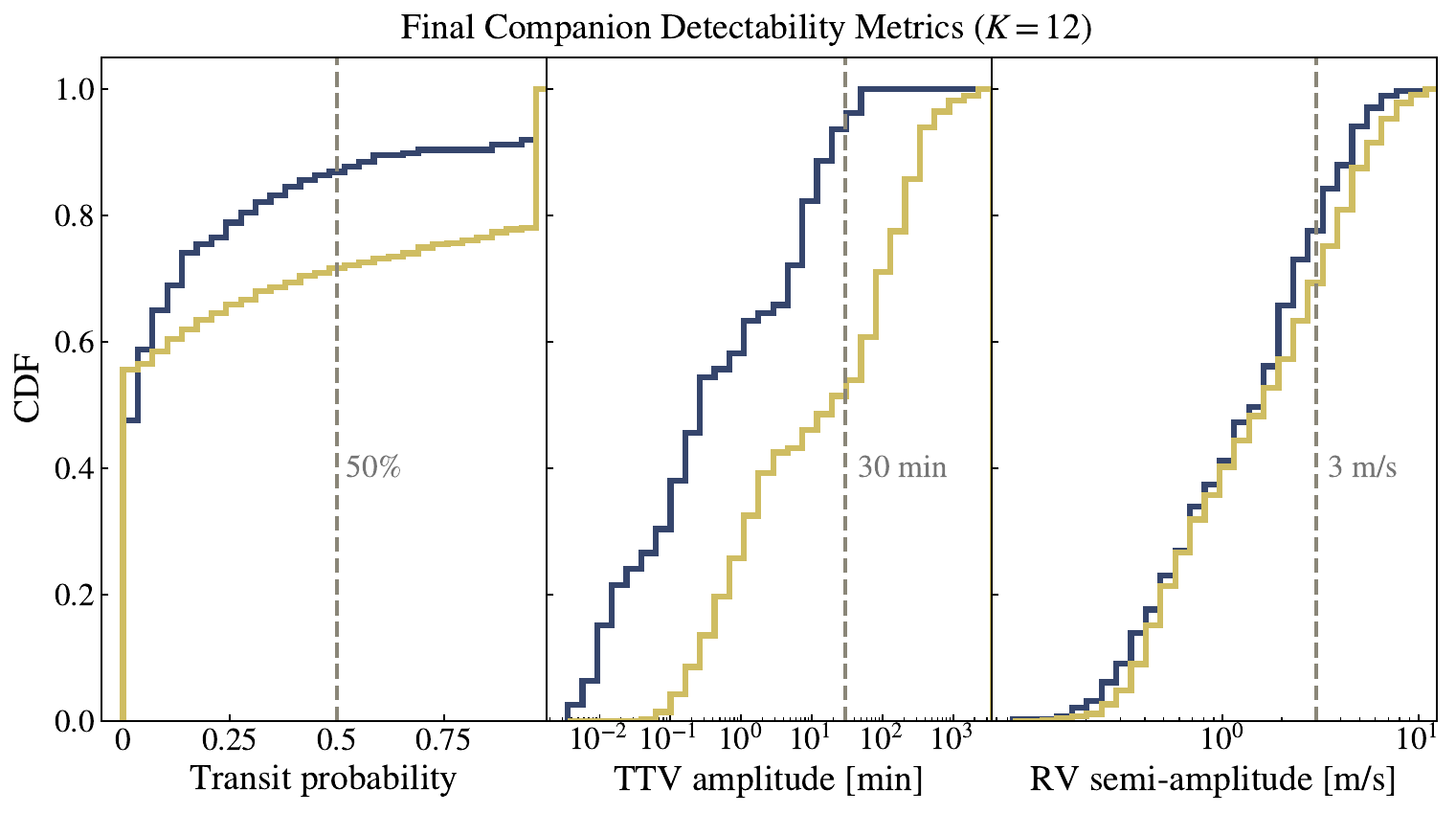}
    } \\ 
    \subfloat[CDFs for our $K=10$ + tides simulations. See panel (a) above for plot legend.\label{fig:detectability_CDFs_nonfid_c}]{
        \includegraphics[width=0.6\linewidth]{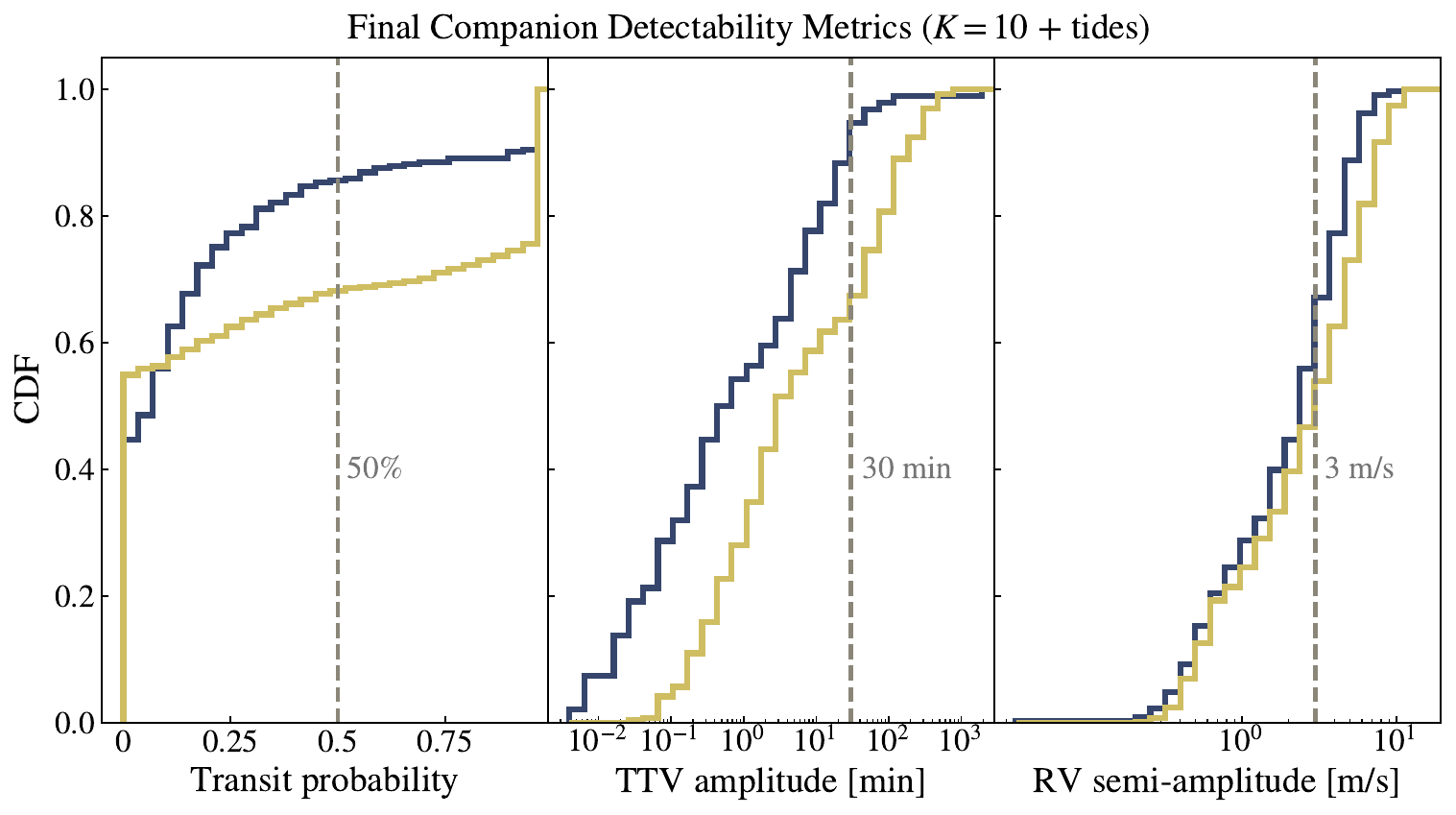}
    }
    \caption{Companion transit probability (left), total TTV amplitude (middle), and RV semi-amplitude (right) CDFs for our non-fiducial simulation suites: (a) $K=8$ results, (b)  $K=12$ results, and (c) $K=10$ + tides results. See caption of Figure \ref{fig:detectability_CDFs} for further plot details. Hot Jupiters' companions are more difficult to detect by each metric, across all simulation suites.}
    \label{fig:detectability_CDFs_nonfid}
\end{figure*}

Carrying out the same observational analysis performed in Section \ref{subsec:obs_dichotomy}, we display in Figure \ref{fig:detectability_CDFs_nonfid} the CDFs of the companion transit probabilities, TTV amplitudes, and RV semi-amplitudes separately for our hot and warm Jupiters in each of these non-fiducial suites. These CDFs, which our AD tests again show are statistically independent (all satisfying $p\leq2\cdot10^{-3}$), illustrate that hot Jupiters' observed preferential isolation is robust to the changes implemented in our non-fiducial suites. Applying our detection thresholds for each observational method, we also find that a significantly lower fraction of hot Jupiters' companions are detected and summarize these statistics in Table \ref{tab:nonfidresults_summary}. Although our primary conclusions remain unchanged, we highlight next some of the differences seen in these non-fiducial simulations. 

\begin{deluxetable}{cccc}
\tabletypesize{\scriptsize}
\tablecaption{Summary of Key Companionship Results for Non-fiducial Simulations}\label{tab:nonfidresults_summary}
\tablehead{\colhead{Metric$^a$} &
\colhead{Hot Jupiters} & \colhead{Warm Jupiters} & \colhead{$|z\mathrm{-score}|$}}
\tablewidth{300pt}
\startdata
\multicolumn{3}{l}{Period Ratio}\\ \\
\multicolumn{3}{l}{$K=8$}\\
    median($P_\mathrm{out}/P_\mathrm{in}$) & $6.8^{+0.9}_{-0.3}$ & $4.8^{+0.3}_{-0.1}$ & $5.8$\\
    mean($P_\mathrm{out}/P_\mathrm{in}$) & $13.8$ & $7.0$ & \nodata \\
    \\
    \multicolumn{3}{l}{$K=12$}\\
    median($P_\mathrm{out}/P_\mathrm{in}$) & $21.8^{+6.4}_{-1.0}$ & $7.8^{+0.7}_{-0.1}$ & $11.4$\\
    mean($P_\mathrm{out}/P_\mathrm{in}$) & $362.6$ & $25.2$ & \nodata \\
    \\
    \multicolumn{3}{l}{$K=10$ + tides}\\
    median($P_\mathrm{out}/P_\mathrm{in}$) & $11.8^{+2.9}_{-0.5}$ & $6.7^{+0.5}_{-0.1}$ & $6.8$\\
    mean($P_\mathrm{out}/P_\mathrm{in}$) & $40.6$ & $13.2$ & \nodata \\\\ \hline
    \multicolumn{3}{l}{Mutual Inclination}\\ \\
    \multicolumn{3}{l}{$K=8$}\\
    median($\mathcal{I}_\mathrm{mut}$) & $4.1^{+0.3}_{-0.2}$$\degree$ & $3.1^{+0.2}_{-0.1}$$\degree$ & $4.8$
    \\
    mean($\mathcal{I}_\mathrm{mut}$) & $5.5\degree$ & $4.2\degree$ & \nodata
    \\\\
    \multicolumn{3}{l}{$K=12$}\\
    median($\mathcal{I}_\mathrm{mut}$) & $4.3^{+0.5}_{-0.2}$$\degree$ & $2.8^{+0.2}_{-0.1}$$\degree$ & $6.3$
    \\
    mean($\mathcal{I}_\mathrm{mut}$) & $7.9\degree$ & $4.9\degree$ & \nodata
    \\\\
    \multicolumn{3}{l}{$K=10$ + tides}\\
    median($\mathcal{I}_\mathrm{mut}$) & $4.5^{+0.5}_{-0.2}$$\degree$ & $3.7^{+0.2}_{-0.1}$$\degree$ & $3.2$
    \\
    mean($\mathcal{I}_\mathrm{mut}$) & $7.2\degree$ & $5.3\degree$ & \nodata
    \\\\ \hline
    \multicolumn{3}{l}{Detectability}\\ \\
\multicolumn{3}{l}{$K=8$}\\
    $f(p_\mathrm{T}>0.5)$ & $26.2\pm2.5\%$ & $47.4\pm2.0\%$ & $6.7$ \\
    $f(V_\mathrm{TTV}>30\,\mathrm{min})$ & $15.0\pm3.3\%$ & $21.3\pm2.6\%$ & 1.5 \\
    $f(K_\mathrm{RV}>3\,\mathrm{m/s})$ & $57.8\pm2.8\%$ & $66.2\pm1.9\%$ & $2.5$ \\\\
\multicolumn{3}{l}{$K=12$}\\
    $f(p_\mathrm{T}>0.5)$ & $13.4\pm1.8\%$ & $28.8\pm1.2\%$ & $7.2$ \\
    $f(V_\mathrm{TTV}>30\,\mathrm{min})$ & $6.3\pm2.7\%$ & $48.6\pm3.0\%$ & $10.5$ \\
    $f(K_\mathrm{RV}>3\,\mathrm{m/s})$ & $24.3\pm2.2\%$ & $33.4\pm1.3\%$ & $3.6$ \\\\
\multicolumn{3}{l}{$K=10$ + tides}\\
    $f(p_\mathrm{T}>0.5)$ & $14.4\pm2.0\%$ & $32.1\pm1.6\%$ & $7.0$ \\
    $f(V_\mathrm{TTV}>30\,\mathrm{min})$ & $11.6\pm3.3\%$ & $36.2\pm3.0\%$ & $5.6$ \\
    $f(K_\mathrm{RV}>3\,\mathrm{m/s})$ & $43.1\pm2.8\%$ & $53.3\pm1.7\%$ & $3.2$ \\\\
\enddata
\tablenotetext{a}{Reported uncertainties are standard asymmetric errors calculated using the 84th and 16th percentile values, normalized by the number of companions.}
\end{deluxetable} 
\textbf{The Effect of Initial Spacing.} Comparing quantitatively the results our fiducial $K=10$ suite with our non-fiducial $K=8$ and $K=12$ suites, we identify several notable trends and verify that each are upheld when the same cut-off time is adopted for all simulations (i.e.,  $10\,\mathrm{Myr}$). Principally, we see a positive correlation between the relative strength (i.e., $|z\mathrm{-score}|$) of the dichotomy for \textit{each} parameter investigated ($P_\mathrm{out}/P_\mathrm{in},\mathcal{I}_\mathrm{mut},p_\mathrm{T},V_\mathrm{TTV},K_\mathrm{RV}$) and the spacing parameter $K$ (see Table \ref{tab:nonfidresults_summary}). In other words, wider primordial planet-planet spacings appear to enhance hot Jupiters' preferential isolation relative to warm Jupiters. Our period ratio and TTV detection fraction metrics appear to be most sensitive to $K$, exhibiting $|z\mathrm{-score}|$ increases of $\sim2\times$ and nearly $\sim10\times$, respectively from $K=8$ to $K=12$. The former can be explained in part by the inherent isolation effect discussed in Section \ref{subsubsec:inherent_vs_evolved}: a higher system compactness means that a larger fraction of our middle-orbiting giants are packed inside the $P_\mathrm{giant}<10\,\mathrm{days}$ hot Jupiter cut. For our $K=8$ suite (most compact), this cut nearly splits our systems symmetrically, allowing giants positioned out to $j<6$ to be classified as hot Jupiters (compared to just $j<4$ for our $K=12$ simulations), thereby balancing out the initial period ratio bias between the two giant populations. The latter is consistent with the known correlation between TTV amplitude and orbital period (e.g., \citealt{Wu_2023}): our simulated giant planets reach longer orbital periods for simulation suites with wider initial spacings (see spacing prescription in Section \ref{sec:simulation_setup}). Therefore, as $K$ increases, the typical orbital period of our warm Jupiters also increases, inflating their expected TTV amplitudes and enhancing the TTV dichotomy. 

Additionally, we find that larger $K$ values lead to higher companion retention: the mean multiplicity of hot Jupiter (warm Jupiter) companions increases from 3.7 (3.7) at $K=8$ to 5.6 (6.0) at $K=12$. Simultaneously, we also note that the overall detectability of our giants' companions decreases with $K$. Specifically, for both hot and warm Jupiters, the transit, TTV, and RV companion detection fractions each drop by $\sim2\times$ from $K=8$ to $K=12$ (see Table \ref{tab:nonfidresults_summary}). Thus, wider initial spacings may allow for more planets to survive around short-period giants, but such companions are on average more difficult to detect. 
We caution, however, against over-interpretation of any trends with initial spacing due to the heterogeneous dynamical ages achieved across these suites: as $K$ increases, the typical dynamical age decreases since the orbital period distribution for giant planets skews to longer values (i.e., $t_f$ corresponds to fewer multiples of $P_\mathrm{giant}$).

\textbf{The Effect of Tidal Forces.} We find that the inclusion of tidal forces does not have a significant impact on our companion properties, with all median and $|z\mathrm{-score}|$ values from our fiducial suite (reported in Table \ref{tab:results_summary}) being statistically consistent (within $\leq2\sigma$) with those from our $K=10$ + tides suite, except for the mutual inclinations of hot Jupiters' companions (see Table \ref{tab:nonfidresults_summary}). Although still indicative of a significant enhancement relative to warm Jupiters' companions, we see that their $\mathrm{median(}\mathcal{I}_\mathrm{mut}\mathrm{)}$ in our $K=10$ + tides suite is $\approx17\%$ lower than in the fiducial $K=10$ suite, and the associated $|z\mathrm{-score}|$ is $2\times$ lower. This modest decrease is consistent with the long-term damping effects of tidal dissipation, which generally reduces the probability of late-time instabilities --- especially ejections (see Section \ref{sec:dynamics}) --- and erodes residual eccentricities and inclinations. To conduct a fair comparison between the hot Jupiter systems in the tidal and fiducial suites, we examine the fiducial simulations at the non-fiducial cut-off time of $t_f=10\,\mathrm{Myr}$ (or $10^9P_\mathrm{giant}$). Indeed, we find that fewer late-time ejections occur in the tidal suite compared to the fiducial suite (e.g., 30\% fewer over $0.5$--$10\,\mathrm{Myr}$), and that hot Jupiter companion eccentricities are also systematically lower in the tides suite (e.g., the $\mathrm{median(}e\mathrm)$ is 15\% smaller).

One may speculate about the possible effects of tides over timescales longer than what is probed by these simulations. For our non-fiducial $K=10$ + tides suite, we compute the tidal dissipation timescales of our giant planets and their companions at the end of the simulation assuming a wide array of potentially physical effective quality factors $Q/k_2$ (\citealt{Goldreich1963,Goldreich1966,Jackson2008}). Assuming $Q/k_2=[10^4,10^5,10^6]$ for our giants, we find that just $\approx$30\%, 15\%, and 10\%, respectively have dissipation timescales $<1\,\mathrm{Gyr}$. For our super-Earth and sub-Neptune-like companions, we adopt $Q/k_2=\{10^2,10^3,10^4\}$ and derive somewhat larger fractions of $\approx37\%$, 30\%, and 18\%, respectively. We repeat these calculations for our fiducial $K=10$ suite and find comparable results (to within $\approx1\%$), indicating that the longer-term effects of tides may only be important for only a small subset of companions in our systems, and an even smaller fraction of giant planets.   

\bibliography{hj_isolation_not_unique}{}
\bibliographystyle{aasjournalv7}

\end{document}